\documentclass[fleqn,usenatbib]{mnras}

\usepackage{newtxtext,newtxmath}

\usepackage[T1]{fontenc}

\DeclareRobustCommand{\VAN}[3]{#2}
\let\VANthebibliography\thebibliography
\def\thebibliography{\DeclareRobustCommand{\VAN}[3]{##3}\VANthebibliography}

\usepackage{graphicx}	
\usepackage{amsmath}	
\usepackage{tabularx}
\usepackage{physics}
\usepackage{float}
\usepackage{orcidlink}
\usepackage[version=4]{mhchem}
\usepackage{lscape}
\usepackage{rotating}
\usepackage{booktabs}
\usepackage{afterpage}
\usepackage{caption}
\usepackage{float}

\newcommand{\LCDM}{$\Lambda$CDM}
\newcommand{\gravsphere}{\texttt{GravSphere}}

\newcommand{\kms}{kms$^{-1}$}

\newcommand{\vI}{-374.3 $\pm$ 1.7}
\newcommand{\vIII}{-343.4 $\pm$ 1.9}
\newcommand{\vV}{-396.3 $\pm$ 1.5}
\newcommand{\vVII}{-306.8 $\pm$ 1.7}
\newcommand{\vIX}{211.5 $\pm$ 2.6}
\newcommand{\vXXXI}{-198.1 $\pm$ 1.2}
\newcommand{\vXXXII}{-369.8 $\pm$ 0.7}

\newcommand{\vdI}{10.2 $\substack{+1.5 \\ -1.2}$}
\newcommand{\vdIII}{10.7 $\substack{+1.7 \\ -1.4}$}
\newcommand{\vdV}{10.7 $\substack{+1.3 \\ -1.1}$}
\newcommand{\vdVII}{12.2 $\substack{+1.4 \\ -1.2}$}
\newcommand{\vdIX}{10.2 $\substack{+2.5 \\ -1.9}$}
\newcommand{\vdXXXI}{12.1 $\substack{+1.0 \\ -0.9}$}
\newcommand{\vdXXXII}{9.2 $\substack{+0.6 \\ -0.5}$}

\newcommand{\DMI}{0.3 $\substack{+0.6 \\ -0.2}$ }
\newcommand{\DMIII}{1.4 $\substack{+0.7 \\ -0.6}$}
\newcommand{\DMV}{1.7 $\substack{+0.4 \\ -0.5}$}

\newcommand{\DMVII}{0.5 $\substack{+0.6 \\ -0.3}$}
\newcommand{\DMIX}{0.8 $\substack{+0.6 \\ -0.4}$}
\newcommand{\DMXXV}{0.2 $\substack{+0.3 \\ -0.1}$}
\newcommand{\DMXXXI}{0.5 $\substack{+0.3 \\ -0.2}$}
\newcommand{\DMXXXII}{0.3 $\substack{+0.2 \\ -0.1}$}

\newcolumntype{Y}{>{\centering\arraybackslash}X}

\title[It's Not Just Star Formation]{It's Not Just Star Formation: A trend of low dark matter densities in the Andromeda dwarf galaxy system}

\author[C.S. Pickett et al.]{Connor S. Pickett\orcidlink{0000-0002-0684-4277}$^{1}$\thanks{E-mail: c.pickett@surrey.ac.uk},
Michelle L. M. Collins\orcidlink{0000-0002-1693-3265}$^{1}$,
Justin I. Read\orcidlink{0000-0002-1164-9302}$^{1}$,
R. Michael Rich\orcidlink{0000-0003-0427-8387}$^{2}$,
Emily J. E. Charles\orcidlink{0000-0002-5886-4202}$^{1}$,
\newauthor
Erik Tollerud\orcidlink{0000-0002-9599-310X}$^{3}$,
Nicolas Martin\orcidlink{0000-0002-1349-202X}$^{4,5}$,
Scott Chapman\orcidlink{0000-0002-8487-3153}$^{6,7}$,
Alan McConnachie\orcidlink{0000-0003-4666-6564}$^{8}$,
Alessandro Savino\orcidlink{0000-0002-1445-4877}$^{9}$,
\newauthor
Daniel R. Weisz\orcidlink{0000-0002-6442-6030}$^{9}$
\\
$^{1}$Department of Astronomy, University of Surrey, Stag Hill Campus, Guildford GU2 7XH, UK \\
$^{2}$ Department of Physics and Astronomy, University of California, Los Angeles, PAB, 430 Portola Plaza, LA, CA 90095-1547, USA \\
$^{3}$ The Space Telescope Institute, 3700 San Martin Drive, Baltimore, MD 21218, USA\\
$^{4}$Universit\'e de Strasbourg, CNRS, Observatoire astronomique de Strasbourg, UMR 7550, F-67000 Strasbourg, France \\
$^{5}$Max-Planck-Institut f\"{u}r Astronomie, K\"{o}nigstuhl 17, D-69117 Heidelberg, Germany \\
$^{6}$Department of Physics and Atmospheric Science, Dalhousie University, 1453 Lord Dalhousie Drive, Halifax, NS B3H 4R2, Canada \\
$^{7}$Department of Physics and Astronomy,  University of British Columbia, Vancouver, BC, V6T1Z1, Canada \\
$^{8}$NRC Herzberg Astronomy and Astrophysics, 5071 West Saanich Road, Victoria, BC V9E 2E7, Canada\\
$^{9}$Department of Astronomy, University of California Berkeley, Berkeley, CA 94720, USA \\
}

\date{Accepted XXX. Received YYY; in original form ZZZ}

\pubyear{\the\year{2026}}

\begin{document}
\label{firstpage}
\pagerange{\pageref{firstpage}--\pageref{lastpage}}
\maketitle

\begin{abstract}
Dynamical mass modeling of Andromeda (M31) dwarf spheroidal (dSph) galaxies has revealed a growing trend of lower central dark matter (DM) densities than predicted by pure DM structure formation in Lambda Cold Dark Matter (\LCDM) cosmology simulations and lower than most Milky Way (MW) satellites. So far, however, only four of the 35 confirmed M31 dSphs have been successfully mass modeled. In this second paper of a series, we aim to better understand growing Local Group (LG) dSph patterns by mass modeling seven more M31 dSphs: Andromeda I, III, V, VII, IX, XXXI, and XXXII. We update the kinematics of each dwarf and estimate their central dark matter densities at 150 pc using the dynamical Jeans modeling tool, \gravsphere. We also update their DM halo mass, $M_{\rm{200}}$, via abundance matching. We find Andromeda III and V to have central DM densities in line with \LCDM ~expectations, resembling dSphs around the Milky Way. The remaining five dwarfs have anomalously low central densities, continuing a growing trend seen for M31 satellites. We investigate each dwarf's star formation history and find that star formation-induced `DM heating' is disfavored as the sole explanation of these lower central densities. We consider the effect of tides and halo concentration scatter on these systems and predict that they should be on more plunging orbits than their denser counterparts. If this prediction is misaligned with the data, it could necessitate new physics beyond the Standard Cosmological Model.

\end{abstract}




\section{Introduction}

Near-field cosmology has long struggled with tensions between observations of the lowest mass galaxies and their predicted properties from cosmological simulations. One of these tensions, known as the `cusp-core' problem \citep{Flores_1994, Moore_1994, deBlok_2010}, is born from a mismatch between observations and \LCDM ~dark matter-only simulations of the dynamics of stars in the center of these galaxies. The former predicted that dwarf spheroidal galaxies should have central high-density DM `cusps,' with diverging densities of $\rho \propto r^{-1}$ \citep{Dubinski_1991, Crone_1994, Navarro_1996a}. However, observations showed an abundance of these systems to have central low- or constant-density `cores' \citep{Flores_1994, Moore_1999}. This contrast in results called to question the \LCDM ~model, with alternative models being predicted such as Self-Interacting Dark Matter (SIDM; \citealt{Spergel_2000}), Warm Dark Matter (WDM; \citealt{Bode_2001}),and Fermionic Dark Matter \citep{Arguelles_2023}. It was also considered that dark matter could be affected by baryonic matter, primarily through processes like `dark matter heating'. Processes of constant gas inflow/outflow, such as supernovae or violent star formation, could `heat' dark matter particles to different orbits and irreversibly cause inner DM halos to expand \citep{Read_2005, Pontzen_2012, DiCintio_2014, Pontzen_2014, Read_2016}. Over time, simulations shifted from DM-only to include baryonic matter. In a \LCDM ~paradigm, simulated dark matter heating scenarios appeared to create cores similar to those observed in dSphs today. However, as shown by \citet{Read_2005} and \citet{Pontzen_2012, Pontzen_2014}, this heating would need to be driven by persistent bursts of star formation.

Milky Way dSphs have historically been used as testing grounds for many of these tension alleviations, due to their proximity to the Milky Way itself and depth of available data (e.g. \citealt{Koposov_2008}). Higher-central density MW dwarf spheroidals generally agree with what is simulated. For example, Draco is estimated to have a high central DM density and relatively short star formation history (e.g. \citealt{DiCintio_2014, Weisz_2014b, Read_2018}). MW satellites with lower central DM densities can typically be reconciled with their extended star formation histories (SFHs) after the epoch of reionization \citep{Grebel_2001, Weisz_2014b}, contributing support to the theory of dark matter heating. One metric for measuring the efficacy of dark matter heating is the relation between stellar mass and halo mass, $M_{*}$/$M_{\rm{200}}$ \citep{Penarrubia_2010, DiCintio_2014}. \citet{Read_2016} and \citet{Read_2019} find that star formation is only effective in transforming cusps to cores in MW satellites in systems above a $M_{*}$/$M_{\rm{200}}$ threshold of $\sim 5 \times 10^{-4}$ \citep{DiCintio_2014}. These findings, from simulated data informed by Milky Way systems (e.g. Making Galaxies in a Cosmological Context (MaGICC); \citealt{Brook_2014}), confirmed that supernova feedback could result in cored DM profiles due to halo expansion. They also found that, above a certain mass threshold ($M_{\rm{200}} > 10^{10.8} \rm{M}_{\rm{\odot}}$), halos could suppress SN-driven outflows and retain their DM cusps. This trend is shown well-exemplified by the Sculptor dwarf galaxy, which resides in an extended DM halo of mass $M_{\rm{200}} \approx 2.0 \times 10^{9}$ M$_{\rm{\odot}}$ \citep{Brook_2015, Read_2019}. The dwarf is estimated to have a central DM density profile that appears intermediate between cusped and cored, with studies in the literature straddling both possibilities (e.g. \citealt{Walker_2011, Amorisco_2011, Strigari_2018}). Sculptor has also been shown to quenched relativity early in its lifetime. Systems in less massive DM halos, such as the Draco dwarf galaxy ($M_{\rm{200}} \approx 1.8 \times 10^{9}$ M$_{\rm{\odot}}$), have similar quenching times as Sculptor and retain their dark matter cusps \citep{Read_2019}. Conversely, more extended star formation within dwarfs of higher-mass, more extended halos (i.e. Fornax of halo mass $M_{\rm{200}} \approx 2.2 \times 10^{10}$ M$_{\rm{\odot}}$) are likely to have dark matter cores \citep{Brook_2015, Read_2019, Rusakov_2021}. Whereas the densest halos should commonly form more stars \citep{Boylan_2011, Boylan_2012, Read_2016}, there are also MW dwarfs inhabiting seemingly extreme low-density halos (i.e. Crater II, Antlia II) which have had more extended star formation than expected \citep{Walker_2009, Collins_2011, Torrealba_2016, Torrealba_2019}. These systems appear anomalously around the MW and tend to be on eccentric orbits that bring them within 30 kpc of the Galactic center (e.g. \citealt{Kirby_2013, Fu_2019, Ji_2021}). This implies that their low densities are a result of tidal forces, not DM heating via star formation, that may have stripped 90\% or more of their matter \citep{Sanders_2018, Fu_2019, Torrealba_2019, Sameie_2020}. Despite some outliers, MW dwarfs align with the idea that denser halos should be host to more star formation and that the DM heating from these SFHs can indeed lower the central DM density of dSphs. \citet{Muni_2025} has also shown that, while there is scatter present in dwarf galaxy stellar mass-halo mass relations, this tends to be due to assembly history and remains consistent with what is observed in MW satellites.

Conversely, upon initial inspection, Andromeda dSphs appear to be inconsistent with dark matter heating processes and tidal interactions seen for MW dwarfs. Through recent modeling of M31 dwarfs, most appear in the low-density regime, lower than is expected in the Standard Cosmology and consistent with DM cores \citep{Collins_2021, Charles_2022, Pickett_2025}. Andromeda VI is the only system, to date, that has a high central density, consistent with a DM cusp. A key issue lies in the star formation histories of M31 satellites, which appear have prominent bursts of star formation shortly after reionization and intermediate quenching times \citep{Weisz_2019, Savino_2023, Savino_2025}. Moreover, Andromeda satellite galaxies appear to have quenched around the same time, at ${\sim}5-6$ Gyr ago \citep{Savino_2025}. This behavior is not typically seen in MW dwarfs of comparable size and luminosity, most of which appear to have `normal' dark matter densities \citep{Read_2019, Pickett_2025}. Brighter satellites of both hosts have more extended star formation, though this is exaggerated for dwarfs at a greater present-day distance from M31 \citep{Savino_2025}. There also appears to be a trend where more isolated M31 dSphs appear to host higher central DM densities compared to more closely-orbiting dwarfs \citep{Collins_2021, Charles_2022, Pickett_2025}. The peculiar star formation patterns around M31, as well as their reliance on distance, may hint at different physical processes affecting its dwarfs' DM densities over their lifetimes. One such reason may be the elevated merger history of Andromeda compared to the MW \citep{McConnachie_2018}. Alternatively, these trends could also be explained by: (i) tidal interactions with Andromeda itself (e.g. \citet{Charles_2022}, Kim et al., in prep.), which is two-fold. One scenario involves dark matter being stripped away during a close orbit to its host. The other scenario involves the shocking and subsequent heating of dark matter as a result of constant expansion and contraction of matter. Accurately understanding this scenario requires understanding the orbit of a dwarf through its proper motion (PM). While there are a handful of PMs now available for M31 systems \citep{Casetti_2024, Casetti_2025}, they are extremely limited and time-consuming (on the scale of decades) to measure. (ii) Low-concentration DM halos (e.g. \citealt{Kim_2024, Julio_2024}), where a fixed halo mass, $M_{\rm{200}}$, can allow for more rapid star formation before reionization and early quenching times. This can lead to a higher stellar mass within a low-density halo, one that does not need to be explained via star formation at all. Other approaches, such as exotic dark matter, could explain these lower concentrations that are not easily recreated in a \LCDM ~paradigm.

With an increasing number of mass modeled dwarfs, trends in the Local Group have become more evident. One such trend lies in the relation between DM density, $\rho_{\rm{DM}}$, and half-light radius, $r_{\rm{h}}$. This relation, discussed further in $\S$\ref{sec:discussion}, could potentially indicate a common estimator whether the DM density of a dwarf is inherently linked to the size of its stellar component. More galaxies are needed to understand this $\rho_{\rm{DM}}$-$r_{\rm{h}}$ relation, especially at the lower-mass end of the dwarf galaxy samples. To better understand these relations, a wider spectrum of observed dwarf galaxy size is needed. Upcoming surveys will see deeper and be able to resolve much fainter stellar populations, such as LSST with expected depths of $r \sim 27.5$ magnitude for MW systems \citep{Ivezic_2019}. Surveys like SAGA \cite[e.g.][]{Geha_2017, Geha_2024} and SEAMLESS \cite[e.g.][]{Jones_2023, Fielder_2025, MutluPakdil_2025} are already revealing the stellar properties of ultra-faint systems 3-5 Mpc from the Milky Way. By detecting more candidate LG dwarfs where we can follow-up their dynamics, star formation histories and (eventually) proper motions we may revolutionize our understanding of the smallest end of dwarfs. To do so, we will need modeling methods that can handle extremely low-membership samples. Recent updates to \gravsphere, known as \gravsphere2 \citep{Banares_2025}, have been shown to mass model systems with as low as 10 member stars with an accuracy of at least 68\% confidence intervals. Further analysis using Graph Neural Networks (GNNs) \citep{Nguyen_2023, Nguyen_2025} have also been shown to return similar results as previous versions of \gravsphere ~from simulated low-N data created in the Feedback In Realistic Environments (FIRE-2; \citealt{Hopkins_2018}) suite. These observational and analytical advancements will continue to redefine our understanding of the low-mass end of dwarf galaxies, reducing scatter in the $M_{\rm{*}}$/$M_{\rm{200}}$ relation and helping to determine the dynamical processes at play within the smallest Andromeda dwarfs.

In this paper, following the same method described in \citet{Pickett_2025} (Paper I, hereafter), we aim to analyze seven more Andromeda dwarf galaxies: I (And I), III (And III), V (And V), VII (And VII), IX (And IX), XXXI (And XXXI), and XXXII (And XXXII). These systems were chosen due to their larger number of confirmed dynamical member stars ($\geq$ 20 stars, \citealt{Tollerud_2012, Collins_2013, Martin_2014}). They vary in size, luminosity, and distance to M31, which provides a suitable data set with which to more comprehensively understand the population as a whole. Much of the analysis performed on these systems is heavily influenced by Paper I, with changes to the process described in this publication. We anticipate a wide range of velocities and dark matter densities in comparison to the preceding publication. We continue to use the Jeans modeling program, \gravsphere, to perform similar estimations on all seven dwarfs as with those in \citet{Collins_2021}, \citet{Charles_2022}, and Paper I. We compare the results of our mass modeling to the existing catalog of LG dSphs and further analyze them with respect to their star formation histories.

This paper is organized as follows: In $\S$ 2, we outline the photometric and spectroscopic observations of each dwarf. This
section also describes the selection of member stars from these data.
In $\S$ 3, we briefly describe the mass modeling technique employed by \gravsphere. In $\S$ 4, we describe the stellar membership selection process for each dwarf and their respective mass modeling results. We discuss these results in $\S$ 5, where we also compare these M31 satellites with others in the Local Group. Finally, we conclude this work in $\S$ 6.

\begin{table}
\renewcommand{\arraystretch}{1.1}
    \caption{Keck DEIMOS spectroscopic mask count, mask exposure times, total observed star count, and final star count used in the analysis for each dwarfs in this work. All masks for And III, And XXXI, and And XXXII were individually exposed for 3600 seconds. Both masks for And V were individually exposed for 2250 seconds.}
    \label{tab:exp_times}
    \setlength{\tabcolsep}{8pt}
    \begin{tabular}{l | c | c | c | c }
    \hline 
    \multicolumn{1}{l}{\textbf{Galaxy}} & \multicolumn{1}{c}{N$_{\rm{mask}}$} & \multicolumn{1}{c}{t$_{\rm{exp}}$ (s)} & \multicolumn{1}{c}{N$_{\rm{*}}$ Observed} & \multicolumn{1}{c}{N$_{\rm{*}}$ Final} \\ \hline
    And I & 2 & 4055, 3600  & 477  & 117\\
    And III & 3 & 3600 & 480  & 171\\
    And V & 3 & 2250 & 454  & 121\\
    And VII & 2 & 3000, 1800 & 357  & 118\\
    And IX & 2 & 2700, 2400 & 248  & 91\\
    And XXXI & 2 & 3600 & 177 & 170\\
    And XXXII & 4 & 3600& 379 & 319
    \end{tabular}
\end{table}

\section{Observations} \label{sec:observations}

\begin{figure*}
\includegraphics[width=\textwidth]{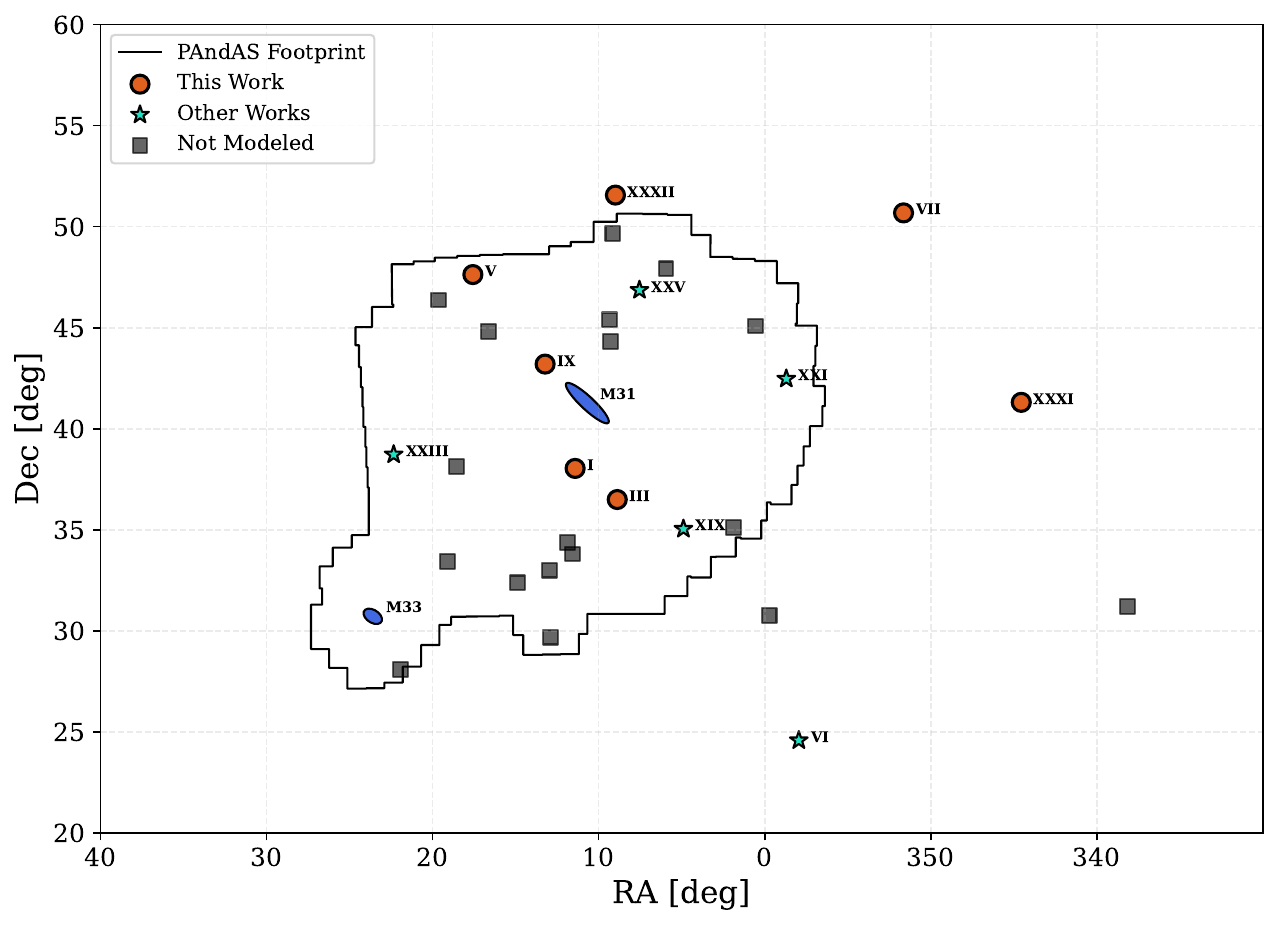}
\centering
\caption{Diagram of the Pan-Andromeda Archaeological Survey footprint \citep{McConnachie_2009, McConnachie_2012}. The solid outline represents the survey area, with M31 shown as an ellipse in the center. M33 is seen as an ellipse at the bottom left of the survey footprint. The dSphs analyzed in this work are shown as thickly-bordered circles and are labeled with their numerical name. dSphs analyzed in other works (And XXI, \citealt{Collins_2021}; And XXV, \citealt{Charles_2022}; And VI/And XXIII, \citealt{Pickett_2025}; And XIX (Pickett et al., in prep)) are shown are as stars and also labeled. M31 dwarfs that have not yet been mass modeled are shown as squares. \textit{Note}: Galaxy sizes are not to scale.} \label{fig:PAndAS_footprint}
\end{figure*}

All dwarfs detailed in this paper were analyzed using existing data. Photometric observations of Andromeda I, III, V, and IX were collected via the Pan-Andromeda Archaeological Survey (PAndAS) from 2008 to 2011 \citep{McConnachie_2009, McConnachie_2012, McConnachie_2018}. Photometry of And VII was collected in August-September 2005 \citep{McConnachie_2005, McConnachie_2007b}, using the Subaru Suprime-Cam, as a follow-up to observations first presented in \citet{McConnachie_2004}. The 3D-projected distance of each dwarf relative to M31 can be seen in Table \ref{tab:dwarf_params}, with their on-sky location relative to the PAndAS survey shown in Figure \ref{fig:PAndAS_footprint}. Spectroscopic data were collected with the Deep Extragalactic Imaging Multi-Object Spectrograph (DEIMOS) on the Keck II Telescope in Hawaii \citep{Faber_2003, McConnachie_2012} between 2005 and 2020 \citep{Collins_2011, Tollerud_2012, McConnachie_2018, Charles_2022}.

And XXXI (Lacerta I/Lac I) and And XXXII (Cassiopeia III/Cas III) were first identified in 2013 by the 3$\pi$ Panoramic Survey Telescope and Rapid Response System 1 (Pan-STARRS1) survey \citep{Martin_2013}. Follow-up DEIMOS spectroscopy was taken of both satellites in 2013, confirming their dwarf status and their bound orbits to M31 \citep{Martin_2014}. The observed parameters for And XXXI and XXXII can be seen in Table \ref{tab:dwarf_params}.

As multiple photometric masks were taken for all dwarfs, it was necessary to remove duplicate stars to maintain cleaner data. This was completed by tagging stars of both the same right ascension and declination, with the star of lower velocity error being kept. If necessary, stars within the data that had velocity error measurements of 0 were also removed, as they were most likely false detections. In doing so, we were able to clean a significant portion of duplicates. The final number of stars used in our analysis can be seen Table \ref{tab:exp_times}. 

\subsection{PAndAS Photometry} \label{subsec:pandas} 
As described in Paper I, PAndAS data were collected between 2008 and 2011 using Mega-Cam on the Canada-France-Hawaii Telescope on Mauna Kea \citep{McConnachie_2009, McConnachie_2012, McConnachie_2018}. The 400 deg$^{2}$ view of M31 halo allowed observations to be taken out to 150 kpc Andromeda's center. This was done with a signal-to-noise ratio (SNR) of $\sim$10 at magnitudes of $g = 25.5$ and $i = 24.5$ \citep{McConnachie_2009}. This magnitude range was ideal for identifying the red giant branch (RGB) stellar population typical of dwarf galaxies, which were classed as old faint stellar overdensities in the PAndAS view. Data were reduced using the Cambridge Astronomical Survey Unit (CASU) pipeline described in \citet{Irwin_2004} \citep{Ibata_2014, McConnachie_2018}. More detail on this reduction can be seen in $\S$2.1 in paper 1. Parameters of PAndAS-observed dwarfs can be seen in Table \ref{tab:dwarf_params}.

\subsection{Subaru Suprime-Cam Photometry} \label{subsec:Subaru} 
Original photometry of And VII was collected using the Isaac Newton Telescope Wide Field Camera (INT WFC) \citep{McConnachie_2004, McConnachie_2005}. INT WFC observations were then followed up on 3-5 August 2005 using the Subaru Suprime-Cam \citep{McConnachie_2007b}. These data were collected in a similar fashion as Andromeda VI and Andromeda XXIII in Paper I and were taken in the Johnson-Cousins $V$- and $I_{\rm{c}}$-bands, with a seeing of 0.5 arcsec \citep{McConnachie_2007a}. These data were also reduced using the CASU pipeline.

\subsection{3\texorpdfstring{$\pi$} ~~Pan-STARRS1 Photometry} \label{subsec:panstarrs}
The Pan-STARRS1 sky survey, located at the Haleakala Observatory in Hawaii, observed the sky north of declination of -30$^{\circ}$ from May 2020 to March 2014. The observatory utilized a 1.4Gpixel, 3.3$^{\circ}$-field-of-view (FoV) camera, with a 1.8m aperture of $f$/4.4 \citep{Martin_2013}. A 5 near-infrared filter system system ($g_{\rm{P1}}r_{\rm{P1}}i_{\rm{P1}}z_{\rm{P1}}y_{\rm{P1}}$) was used, with an 8 $\times$ 8 orthogonal transfer array of CCDs. These arrays were subdivided into 8 $\times$ 8 `cells', each with an independent 590 $\times$ 598 10 $\mu$m pixel CCD \citep{Tonry_2012}. Rapid sky mapping, with each survey location being observed four times per year, allowed for a median seeing value in respective $g_{\rm{P1}}r_{\rm{P1}}i_{\rm{P1}}z_{\rm{P1}}y_{\rm{P1}}$ filters of 1.27, 1.16, 1.11, 1.06, and 1.01 arcseconds. Stacked data of Pan-STARRS1 were created via processing through the Image Processing Pipeline described in \citet{Magnier_2006}, \citet{Sterken_2007}, and \citet{Magnier_2008}.

The Andromeda region had been observed for three consecutive seasons at the time of discovery of And XXXI and XXXII \citet{Martin_2013}. This resulted in a respective 10$\sigma$-completeness $r_{\rm{P1}}$- and $i_{\rm{P1}}$-band limit of 22.1 and 22.2 for And XXXI, and a 22.0 and 21.9 limit for And XXXII. Both bands yielded an area coverage of $> 95\%$ based off the presence of PS1-observable 2MASS stars. It is further described that `spurious offsets' were present in stacked photometry, calibrated at the 0.01-magnitude level \citep{Schlafly_2012} for both dwarfs. The region of And XXXI yielded no offset in the $i_{\rm{P1}}$-band but presented a $\sim$0.02 mag offset for $\sim$20$\%$ of sources in the $r_{\rm{P1}}$-band. Likewise, the region around And XXXII was affected by 0.05 mag in the $r_{\rm{P1}}$-band and 0.01 mag in the $i_{\rm{P1}}$-band. The offset data were kept within the catalog, as they still had `reliable' photometry. All sources were then dereddened, assuming a distance to M31 of 779 $\substack{+19 \\ -18}$ kpc, using the \citet{Schlegel_1998} maps and the \citet{Schlafly_2011} Pan-STARRS1 filter extinction coefficients: $A_{g_{\rm{P1}}} / E(B - V) = 3.172$, $A_{r_{\rm{P1}}} / E(B - V) = 2.271$, and $A_{i_{\rm{P1}}} / E(B - V) = 1.682$.

\subsection{DEIMOS Spectroscopy} \label{subsec:deimos}
Spectroscopic observations of all dwarfs discussed in this work were collected by Keck DEIMOS on Mauna Kea \citep{Faber_2003}. This instrument used a multiobject mode with an FoV of 16 $\times$ 8 arcmin$^{2}$, which translates to $\sim$3.5 kpc $\times$ 1.3 kpc at a distance of Andromeda. This meant that each dwarf we discuss fit within the DEIMOS footprint to a varying degree, with some systems requiring multiple masks. The spectrograph observes in the optical to near-infrared  spectrum ($\lambda = 4000$ \AA ~- $1100$ \AA, within which Calcium II (Ca II) triplet absorption lines can be identified at $\sim8500$ \AA. These features are used to determine velocities and metallicities of observed stars within a given galaxy. Each observation was collected with the same 1200mm$^{-1}$ line grating at a resolution of 1.3 \AA, centered at $\lambda \sim$ 7800 \AA ~to resolve Ca II lines. Each dwarf was observed for varying exposure times, which are shown in Table \ref{tab:exp_times}. The observations are described in \citet{Tollerud_2012} and \citet{Collins_2013} for all dwarfs except And XXXI and And XXXII, which are described in \citet{Martin_2014}. We use the catalog data from each of these surveys without further processing. 

 Data for the dwarfs presented in \citet{Tollerud_2012} were reduced using the \texttt{SPEC2D} DEIMOS reduction pipeline, as well as modified \texttt{SPEC1D} and \texttt{ZSPEC} codes \citep{Cooper_2012, Newman_2013}. Data for dwarfs presented in \citet{Collins_2013} were reduced using a custom pipeline described in \citet{Ibata_2011}. Both reductions follow similar steps and are directly compared in \citet{Collins_2020}, which shows them to be consistent. Both reductions identify and correct for cosmic rays, scattered light, illumination, slit function, and fringing. Pixel-to-pixel variation corrections are made via flat-fielding, while pixel wavelength calibrations are completed through arc-lamp exposures. Prior to spectrum extraction, sky subtractions were taken of a small spatial region around each target from the 2D spectra without resampling. The inherent DEIMOS uncertainty in Ca II velocity measurements remains at 3.2 kms$^{-1}$, consistent with \citet{Collins_2020} and Paper I. All velocities were also corrected to the heliocentric frame. Slitmask misalignments have been noted to cause velocity shifts up to 15 kms$^{-1}$. These are corrected by comparing telluric absorption lines in the data to atmospheric models, shifting each spectrum to the correct frame \citep{Ibata_2011, Collins_2013}. Velocity gradients were not found in any mask. As in Paper I, the total velocity uncertainty of each star is a combination of MCMC posterior distribution uncertainty, inherent DEIMOS uncertainty, and reduction pipeline uncertainty at 3.2 kms$^{-1}$. More details on how this value is used can be seen in $\S$ \ref{subsubsec:CMDCut}.

\thispagestyle{empty}
\begin{sidewaystable*}
    \vspace*{18cm}
    \centering
    \setlength{\tabcolsep}{8pt}
    \renewcommand{\arraystretch}{1.8}
    
    \caption{Properties of the dwarf galaxies analyzed in this work. Sources: $^{a}$\citealp{Martin_2014}, $^{b}$\citealp{Martin_2016}, $^{c}$\citealp{McConnachie_2018}, $^{d}$\citealp{Savino_2022}, $^{e}$\citealp{Charles_2022}, $^{f}$this work. \textit{Note}: The $\rho_{\mathrm{DM}}$ and $M_{\mathrm{200}}$ values for Andromeda XXV have been updated using updated SFH data from \protect\citet{Savino_2025}, with all remaining values taken \protect\citet{Charles_2022}.} 
    
    \vspace*{0.25cm}
    
    \label{tab:dwarf_params}
    \begin{tabular}{l c c c c c c c c c}
        \toprule
            \textbf{Galaxy} & $\alpha$, $\delta$ (J2000) & m$_{\mathrm{v}}$ & M$_{\mathrm{v}}$ & D$_{\mathrm{\odot}}$ (kpc) & D$_{\mathrm{M31}}$ (kpc) & $\epsilon$ & $\theta ~(^{\circ})$ & r$_{\mathrm{h}}$ (") & r$_{\mathrm{h}}$ (pc)\\ \hline
            
            $^{b,c,d}$And I & $0^{\mathrm{h}}45^{\mathrm{m}}39\overset{\mathrm{s}}{.}7 \pm 0\overset{\mathrm{s}}{.}3, +38^{\circ}02'15'' \pm 6''$ & 24.3 $\pm$ 0.05 & -11.4 $\pm$ 0.2 & 721.1 $\substack{+17 \\ -16}$ & 48.0 $\substack{+10 \\ -3.2}$ & 0.6 $\pm$ 0.04 & 140 $\pm$ 3 & 2.2 $\pm$ 0.2 & 420 $\pm$ 43 \\
            
            $^{b,c,d}$And III & $0^{\mathrm{h}}35^{\mathrm{m}}30\overset{\mathrm{s}}{.}9 \pm 0\overset{\mathrm{s}}{.}4, +36^{\circ}29'56'' \pm 8''$ & 24.3 $\pm$ 0.05 & -9.5 $\pm$ 0.2 & 721.1 $\substack{+17 \\ -16}$ & 84.9 $\substack{+19 \\ -14}$ & 0.6 $\pm$ 0.04 & 140 $\pm$ 3 & 2.2 $\pm$ 0.2 & 420 $\pm$ 43 \\
            
            $^{b,d}$And V & $1^{\mathrm{h}}10^{\mathrm{m}}17\overset{\mathrm{s}}{.}5 \pm 0\overset{\mathrm{s}}{.}4, +47^{\circ}37'42'' \pm 6''$ & 24.4 $\pm$ 0.06 & -9.3 $\pm$ 0.2 & 758.6 $\pm$ 21 & 110.5 $\substack{+7 \\ -3.5}$ & 0.26 $\substack{+0.09 \\ -0.07}$ & 54 $\pm$ 10 & 1.4 $\pm$ 0.2 & 353 $\substack{+35 \\ -24}$ \\
            
            $^{c,d}$And VII & $23^{\mathrm{h}}26^{\mathrm{m}}31\overset{\mathrm{s}}{.}7, +50^{\circ}40'33.0''$ & 24.6 $\pm$ 0.06 & -12.8 $\pm$ 0.3 & 824.1 $\pm$ 23 & 230.8 $\substack{+8.4 \\ -6.5}$ & 0.13 $\pm$ 0.04 & 94 $\pm$ 8 & 3.5 $\pm$ 0.1 & 815 $\pm$ 28 \\
            
            $^{b,c,d}$And IX & $0^{\mathrm{h}}52^{\mathrm{m}}53\overset{\mathrm{s}}{.}4, +43^{\circ}11'57'' \pm 8''$ & 24.2 $\pm$ 0.06 & -8.6 $\pm$ 0.3 & 701.5 $\substack{+20 \\ -19}$ & 82 $\substack{+26 \\ -24}$ & 0.0 $\substack{+0.2 \\ -0.0}$ & 41 $\pm$ 65 & 2.5 $\pm$ 0.1 & 408 $\substack{+62 \\ -42}$ \\

            $^{e}$And XXV & $0^{\mathrm{h}}30^{\mathrm{m}}09\overset{\mathrm{s}}{.}9, +46^{\circ}51'41''$ & 15.3 $\substack{+0.3 \\ -0.2}$ & -9.1 $\substack{+0.3 \\ -0.2}$ & 751.6 $\substack{+25 \\ -21}$ & 85.2 $\substack{+12 \\ -4.4}$ & 0.03 $\substack{+0.2 \\ -0.03}$ & -16 $\pm$ 30 & 2.7 $\substack{+0.4 \\ -0.3}$ & 590 $\substack{+90 \\ -47}$ \\
            
            $^{a}$And XXXI (Lac I) & $22^{\mathrm{h}}58^{\mathrm{m}}16\overset{\mathrm{s}}{.}3, +41^{\circ}17'28.0''$ & 24.4 $\pm$ 0.05 & -11.2 $\pm$ 0.3 & 744.7 $\pm$ 17 & 261.4 $\substack{+6.9 \\ -5.9}$ & 0.4 $\pm$ 0.07 & -59 $\pm$ 6 & 4.2 $\substack{+0.4 \\ -0.5}$ & 910 $\substack{+89 \\ -110}$ \\
            
            $^{a}$And XXXII (Cas III) & $0^{\mathrm{h}}35^{\mathrm{m}}59\overset{\mathrm{s}}{.}4, +51^{\circ}33'35''$ & 24.5 $\pm$ 0.06 & -12.4 $\pm$ 0.2 & 801.7 $\pm$ 23 & 146.8 $\substack{+7.8 \\ -4.2}$ & 0.5 $\pm$ 0.09 & -90 $\pm$ 7 & 6.5 $\substack{+1.2 \\ -1.0}$ & 1516 $\substack{+283 \\ -237}$ \\
        \bottomrule
    \end{tabular}

    \vspace{1.0cm}
    
    \renewcommand{\arraystretch}{1.8}
    \setlength{\tabcolsep}{13.25pt}
    \begin{tabular}{l c c c c c c c c c}
        \toprule
        \textbf{Galaxy} &  N$_{\mathrm{mem}}$ & \shortstack{v \\ (kms$^{-1}$)} & \shortstack{$\sigma_{\mathrm{v}}$ \\(kms$^{-1}$)} & \shortstack{$L$ ($r < r_{\mathrm{h}}$) \\ ($\times 10^{5}$ L$_{\mathrm{\odot}}$)} & \shortstack{$M$ ($r < r_{\mathrm{h}}$) \\ ($\times 10^{7}$ M$_{\mathrm{\odot}}$)} & \shortstack{[M/L] \\ (M$_{\mathrm{\odot}}$/L$_{\mathrm{\odot}}$)} & \shortstack{[Fe/H] \\ (dex)} & \shortstack{$\rho_{\mathrm{DM}}$ \\ ($\times 10^{8}$ M$_{\mathrm{\odot}}$ kpc$^{-3}$)} & \shortstack{$M_{\mathrm{200}}$ \\ ($\times 10^{9}$ M$_{\mathrm{\odot}}$)} \\ \hline
        
        $^{f}$And I & 42 & \vI & \vdI & 30.2 $\pm$ 5.6 & 5.3 $\substack{+1.5 \\ -1.3}$ & 17.4 $\substack{+6.0 \\ 5.4}$ & -1.5 & \DMI & 9.3 $\pm$ 2.3\\
        
        $^{f}$And III & 37 & \vIII & \vdIII & 5.2 $\pm$ 1.0 & 2.8 $\substack{+0.9 \\ -0.8}$ & 53.2 $\substack{+20.0 \\ -17.8}$ & -1.9 & \DMIII & 3.0 $\pm$ 1.0 \\
        
        $^{f}$And V & 61 & \vV & \vdV & 4.4 $\pm$ 0.8 & 2.3 $\substack{+0.6 \\ -0.5}$ & 53.3 $\substack{+16.8 \\ -15.0}$ & -1.85 & \DMV & 3.4 $\pm$ 1.9 \\
        
        $^{f}$And VII & 63 &\vVII & \vdVII & 107.7 $\pm$ 29.7 & 7.05 $\substack{+1.6 \\ -1.4}$ & 6.6 $\substack{+2.3 \\ -2.2}$ & -1.32 & \DMVII & 27.0 $\pm$ 3.6\\
    
        $^{f}$And IX & 20 & \vIX & \vdIX & 2.4 $\pm$ 0.7 & 2.5 $\substack{+1.3 \\ -1.0}$ & 104.6 $\substack{+60.3 \\ -50.0}$ & -2.03 & \DMIX & 2.2 $\pm$ 0.8 \\

        $^{e,f}$And XXV & 49 & -107.7 $\pm$ 1.0 & 3.7$\substack{+0.4 \\ -0.5}$ & 3.7 $\substack{+0.4 \\ -0.5}$ & 0.5 $\pm$ 0.3 & 25 $\substack{+17 \\ -16}$ & -1.9 & \DMXXV & 2.8 $\pm$ 0.5 \\
        
        $^{a,f}$And XXXI & 115 & \vXXXI & \vdXXXI & 25.4 $\pm$ 7.0 & 7.8 $\substack{+1.4 \\ -1.5}$ & 30.7 $\pm$ 10.2 & -2.0 & \DMXXXI & 12.3 $\pm$ 5.7 \\
        
        $^{a,f}$And XXXII & 189 & \vXXXII & \vdXXXII & 75.9 $\pm$ 14.0 & 7.2 $\substack{+1.7 \\ -1.4}$ & 9.5 $\substack{+2.8 \\ -2.6}$  & -1.7 & \DMXXXII & 18.3 $\pm$ 3.3 \\
        \bottomrule
    \end{tabular}
\vspace*{\fill}
\end{sidewaystable*}

\section{Mass Modeling Method} \label{sec:massmodeling}

We use the same dynamical mass-modeling tool,\gravsphere, ~described in Paper I for our analysis in this work. The software is described in detail in \citet{Read_2017}, \citet{Read_2018}, \citet{Genina_2020}, and \citet{Collins_2021}. \gravsphere ~solves the spherical Jeans equation for a set of member stars, used as 'tracers,' with radial velocity measurements. Doing so allows for the estimation the dark matter density, $\rho(r)$, and the velocity anisotropy profile, $\beta(r)$. 

We use the \texttt{CoreNFWTides} model described in \citet{Read_2018} and \citet{Collins_2021}, which fits DM density via a cusped Navarro-Frenk-White (NFW) profile \citep{NFW_1996b}. This profile intrinsically contains a virial mass, $M_{\rm{200}}$, and a concentration parameter, $c_{\rm{200}}$. The \texttt{CoreNFWTides} model introduces four new parameters ($n$, $r_{\rm{c}}$, $r_{\rm{t}}$, $\delta$) to better control the shape of the DM density fit. Our prior values for all \texttt{CoreNFWTides} parameters can be seen Table \ref{tab:coreNFWpriors}. A more detailed description each of these parameters, as well as each equation used for fitting, can be found in \citet{Pickett_2025}. The \gravsphere ~priors used for each dwarf can be also be seen in Table \ref{tab:coreNFWpriors}.

\gravsphere ~has been tested on triaxial dwarfs with a similar number of available stellar tracers as our dwarfs in this paper \citep{Read_2017, Read_2018, Genina_2020, Collins_2021, Nguyen_2025}. These tests result in bias smaller than our quoted uncertainties, despite a system's given asymmetry. \citet{deLeo_2024} has also shown that the software works in the presence of extreme tides through rigorous mock data testing. Further safeguards against binning biases, specifically towards cusped profiles, have been implemented for low-member and large velocity error data. This is done via \texttt{Binulator}, a built in binning software that fit a Gaussian probability distribution to each velocity bin, providing stronger estimates of the mean, variance, kurtosis, and uncertainties. These binned results are then fed into \gravsphere. \citet{Charles_2022} shows an example of this process, where only 49 member stars are used for the analysis of Andromeda XXV. More testing results can also be seen in Appendix A of \citet{Collins_2021}.

\section{Analysis and Results} \label{sec:analysis}

\subsection{Membership Probability} \label{subsec:membership}

Membership probability cuts for each dwarf galaxy are determined much the same way as in Paper I. DEIMOS velocity data for a given dwarf are matched to the associated photometry, requiring a maximum separation of 1-2 arcseconds between sources, using the \texttt{astropy} Python package \citep{Astropy_2013, Astropy_2018, Astropy_2022}. A separation of 2 arcseconds is only applied to dwarfs observed with the Spectroscopic and Photometric Landscape of Andromeda's Stellar Halo (SPLASH) dwarfs (And I, III, V, IX; \citealt{Tollerud_2012}) due to differing observatories used for photometric and spectroscopic data collection. As And VII was observed with Subaru, our DEIMOS spectroscopy matched sources within 1 arcsecond. For PAndAS and Subaru photometry, a CMD is constructed with the matched stars, within which an RGB can be identified. This RGB is then fit using a best-fit-by-eye isochrone. Pan-STARRS photometry, however, are too shallow to resolve a dwarf's RGB and can therefore not be analyzed in the same fashion. This difference is further explained below. Regardless of the applied CMD membership technique, the velocity of RGB stars with probability of $P_{\rm{CMD}} > 0.1$ are then used in a Markov Chain Monte Carlo (MCMC) calculation to further estimate their kinematic membership. Each cut is described in further detail below and in full detail in Paper I.

\subsubsection{CMD Probability} \label{subsubsec:CMDCut}

CMD probabilities for PAndAS photometry are estimated by overlaying an old (10-12 Gyr), metal-poor ($-1.5<$[Fe/H]$<-2.1$), alpha enhanced ($[\alpha/$Fe]$=+0.4$) isochrone to the color-magnitude diagram of each system via a \texttt{Padova} model \citep{ODonnell_1994, Kroupa_2001, Kroupa_2002, Groenewegen_2006, Marigo_2008, Marigo_2013,Chen_2014, Tang_2014, Chen_2015, Rosenfield_2016, Pastorelli_2019, Trabucchi_2019, Bohlin_2020, Pastorelli_2020, Trabucchi_2021}. The distance of each catalog-matched star is measured to a best-fit-by-eye isochrone, which is interpolated onto a fine grid using the \texttt{scipy.interpolate} package \citep{Scipy_2020}. The minimum distance between an RGB star and the isochrone is calculated then applied to a Gaussian probability:

\begin{equation} \label{eq:Gauss_CMD}
    P_{\rm{CMD}} = \exp \Bigg(-\frac{1}{2} \Bigg(\frac{R_{\rm{min}}^{2}}{\eta _{\rm{CMD}}^{2}}\Bigg)\Bigg)
\end{equation}
where $\eta_{\rm{CMD}}$ is a scaling parameter that accounts for the scatter of stars around the isochrone. We use a varying value of $\eta_{\rm{CMD}}$ for each dwarf, depending on the depth of the data. This can be seen for each system in Table \ref{tab:prob_priors} in Appendix \ref{sec:append_plots}. Varying this parameter allows for a reasonable width for the RGB as opposed to considering a single stellar population. 

Pan-STARRS photometric data (And XXXI, And XXXII) are instead assigned CMD membership probabilities with a method similar to that of \citet{Martin_2014}. A key difference in our work is that we use Pan-STARRS DR2, while \citet{Martin_2014} uses DR1. We also trim the catalog-matched data to only contain stars within 2$r_{\rm{h}}$ of a given dwarf as opposed to 3$r_{\rm{h}}$. The trimmed data are then used to plot a $g-i$ CMD, within which we draw a quadrilateral around an identified over-density of stars. All contained stars are assigned a probability of $P_{\rm{RGB}} = 1$, while stars outside this region are given a probability of $P_{\rm{RGB}} = 0$. 

\begin{table}
\renewcommand{\arraystretch}{1.25}
    \caption{Isochrone parameters for the dwarfs analyzed in this paper, where $A_{\rm{v}}$ is the extinction coefficient. Andromeda XXXI and Andromeda XXXII were not fit with an isochrone and are therefore excluded from this table.}
    \label{tab:iso_params}
    \begin{tabularx}{\columnwidth}{l | Y | Y | Y}
    \hline 
    \multicolumn{1}{l}{\textbf{Galaxy}} & \multicolumn{1}{c}{$A_{\rm{v}}$} & \multicolumn{1}{c}{Age (Gyr)} & \multicolumn{1}{c}{[Fe/H] (dex)} \\ \hline
    And I & 0.0 & 1.2 & -1.5 \\
    And III & 0.2 & 1.2 & -1.9 \\
    And V & 0.05 & 1.2 & -1.85 \\
    And VII & 2.0 & 1.2 & -1.32 \\
    And IX & 0.25 & 1.2 & -2.03
    \end{tabularx}
\end{table}

\subsubsection{Velocity Probability} \label{subsubsec:VelCut}

An MCMC routine, outlined in \citet{Goodman_2010}, is run for each dwarf galaxy via the \texttt{emcee} Python package \citep{Foreman-Mackey_2013}. As stated in Paper I, it is necessary to use a multi-Gaussian fit on Andromeda satellites due to line-of-sight contamination from the Milky Way and M31. A key difference in the analysis of our new dwarfs compared to previous analyses is the inclusion of a Bayesian Information Criterion (BIC) estimation. This was performed via the\texttt{sklearn.mixture GaussianMixture} Python package \citep{sklearn_2011}. This addition allows for the testing of the optimal Gaussian fits to the velocity histogram for a given galaxy.

The inclusion of a BIC was decided upon following the analyses of Andromeda VI and Andromeda XXIII, which both used a four-Gaussian fit \citep{Pickett_2025}. This method, however, is not always a one-size-fits-all MCMC routine. Systems such as And I and And VII have varying levels of contamination contributions from the MW and M31, meaning a two- or three-Gaussian fit will work better. In contrast to the four-fit routine established in \citet{Gilbert_2006}, many Andromeda dwarfs' velocity distributions are fit with three Gaussians, as minimal contamination from Milky Way stars can be easily combined into a singular fit. This is confirmed by our Gaussian Mixture tests and BIC results. It must be noted, however, that the returned BIC parameters are not used as the input parameters for the MCMC routine. The mean velocity, $v_{\rm{peak}}$, for the dwarf galaxy, the Milky Way, and Andromeda are taken from the literature. This is done, in part, to prioritize more direct comparisons to previous kinematic analyses.

The utilization of a BIC is to minimize the risk of over- or under-fitting data, which are unavoidably inconsistent across differing surveys. While our BIC estimations confirm that most of our data are best fit with a three-Gaussian MCMC routine (e.g. Fig. \ref{fig:3FitBIC}), And I is better fit with four Gaussians. This can be attributed to its position in the M31 stellar halo and proximity to the Giant Andromeda Stream (GSS; \citealt{Ibata_2001, Ibata_2004, McConnachie_2018}), both of which cause contamination. The lack of this contamination in other data makes it unnecessary to fit a fourth curve to the remaining dwarfs in this paper. In contrast, And VII is shown to be best fit with two Gaussians. This can be seen through its estimated BIC in Figure \ref{fig:2FitBIC} and is most likely attributed to low halo contamination due to a large distance from M31. Despite this lower number of fits compared to other data in this work, the resulting systemic velocity and velocity dispersion align well with literature values within 1$\sigma$. 

We perform the same velocity MCMC routine for each galaxy, regardless of the number of Gaussian fits, following an analysis consistent with \citet{Collins_2021, Charles_2022}, and \citet{Pickett_2025}. Each routine is run with 200 walkers for 3000 steps, with a burn-in period of 1550 steps. This cutoff is confirmed via an integrated autocorrelation time, with a convergence of 200 walkers being achieved after 1550 steps \citep{Goodman_2010}. The priors used to inform this routine for each galaxy can be seen in Table \ref{tab:vel_priors} in Appendix \ref{sec:append_plots}. The results for the systemic velocity ($v_{\rm{r}}$) and velocity dispersion ($\sigma_{\rm{v}}$) are intermediate values used to determine membership likelihood. These likelihoods are then used as weights in a second MCMC routine outlined in Section \ref{subsec:kinematics}.

We define the total membership probability of each star as $P_{\rm{mem}} = P_{\rm{CMD}}\times P_{\rm{vel}}$. Consistent with \citet{Charles_2022} (C23, hereafter) and Paper I, any stars of $P_{\rm{mem}} \geq 0.10$ are considered members, with a stricter cut potentially limiting lower-probability members. This is further explained in $\S$ 4.1.2 of Paper I. The identified members for each dSph in the present work can be seen in Table \ref{tab:dwarf_params} under column N$_{\rm{mem}}$.

\subsection{Dwarf Kinematics} \label{subsec:kinematics}

A Gaussian logarithmic likelihood is used to weigh each member stars' probability via a single Gaussian fit (Equation 12 in Paper I). A second MCMC routine is run using 500 walkers over 5000 steps, with a burn-in of 2000 steps. These parameters are consistent with previous analyses and confirmed with a convergence autocorrelation time. Initial values for $v_{\rm{r}}$ and $\sigma_{\rm{v}}$ are taken from various sources of literature to inform the MCMC prior ranges shown in Table \ref{tab:vel_priors}.

As with Andromeda VI and Andromeda XXIII, low numbers of members still result in well-resolved velocity values. All of our results fall within 1$\sigma$ of their accepted literature values. We reference the systemic velocity and velocity values shown in \citet{Tollerud_2012} for Andromeda I, III, V, VII, and IX. We reference \citet{Martin_2013} for Lacerta I and Cassiopeia III. Our results are used to recalculate the mass within the half-light radius using the relation from \citet{Walker_2009} (shown in Paper I as Equation 13). Our velocity results and enclosed mass estimates can be seen in Table \ref{tab:dwarf_params}.

\subsection{Mass Profiles} \label{subsec:massprofiles}

In this section, we continue our mass profile analysis of Andromeda dwarfs and compare them to similar dSphs around the Milky Way and M31. This follows the comparisons shown in \citet{Read_2016}, \citet{Read_2017}, \citet{Read_2018}, \citet{Read_2019}, \citet{Genina_2020}, C21, C23, and Paper I. We only show the resulting \texttt{GravSphere} estimation plots and brief descriptions in this section. A more detailed description of how these profiles are fit can be found in $\S$4.3 of Paper I.

\subsubsection{Using \texttt{Binulator} and \gravsphere}\label{subsubsec:BinAndGrav}

Similar to the modeling done for other M31 dwarf spheroidals, a surface brightness profile (SBP) is created with available photometric data for each system. This is accomplished by taking point sources in these data out to $5 ~\times ~r_{\rm{h}}$ from the measured center of the dwarf and measuring the distance of the all RGB stars within this region to the best-fit-by-eye isochrone. To maximize likelihood, and consistent with C23, $\eta_{\rm{CMD}}$ from Equation \ref{eq:Gauss_CMD} is doubled for these measurements to obtain $\eta_{\rm{CMD, SBP}}$. A membership probability, $P_{\rm{dist}}$, is then calculated for each point within a Plummer sphere around the dwarf, using Eq. 14 in Paper I. The value of $\eta_{\rm{dist}}$ used in this calculation, as well as other parameters used to inform \texttt{Binulator} for each dwarf, can be seen in Table \ref{tab:prob_priors} in Appendix \ref{sec:append_plots}.

Consistent with previous analyses, point sources with a value of $P_{\rm{dist}} > 0.01$ are included in the SBP as members. The velocity of each member is determined via Calcium II triplet fitting of available spectral data. The number of photometric and kinematic tracers for each galaxy are determined via the following equation:

\begin{equation}\label{eq:eff_tracers}
    N_{\rm{eff}} = \sum_{i = 0}^{N_{\rm{mem}}} P_{\rm{mem, i}}
\end{equation}
where N is the sum if the probabilities for each star. The number of effective tracers for each system can been in Table \ref{tab:tracer_vals} in Appendix \ref{sec:append_plots}. The SBP of fainter dwarfs can more easily blend into the background. A cut is applied at 1500 pc to the SBP of And XXXI to avoid fitting any background stars with \texttt{Binulator}, similar to the cuts performed on And VI and And XXIII in Paper I. This cut was not necessary for other dwarfs in this publication.

\gravsphere ~fits the SBP of each dwarf, as well as the radial velocity profile, created by \texttt{Binulator}. Diagnostic plots output by \gravsphere ~can be seen in the Appendix \ref{sec:append_plots} as Figures \ref{fig:And1_GSDiag} to \ref{fig:Cas3_GSDiag}. As noted in Paper I, solar distance measurement to a given dwarf do not affect the final DM estimation. Using measured distance uncertainties may help marginalize better over distance, but this is not applied to any of our analysis. Stellar mass is obtained using the relation from \citet{Simon_2019} assuming a stellar mass-to-light ratio of 2 for old stellar populations, where M$_{*} = 2 \times \mathrm{L}_{\rm{*}}$. Note that, in \gravsphere, we use the luminosity calculated via the absolute magnitude $M_{\rm{V}}$ of each dSph and not the enclosed luminosity, L$(r < r_{\rm{h}})$.

\subsubsection{Dark Matter Densities} \label{subsubsec:DMdensities}

\begin{figure*}
    \includegraphics[width=\textwidth]{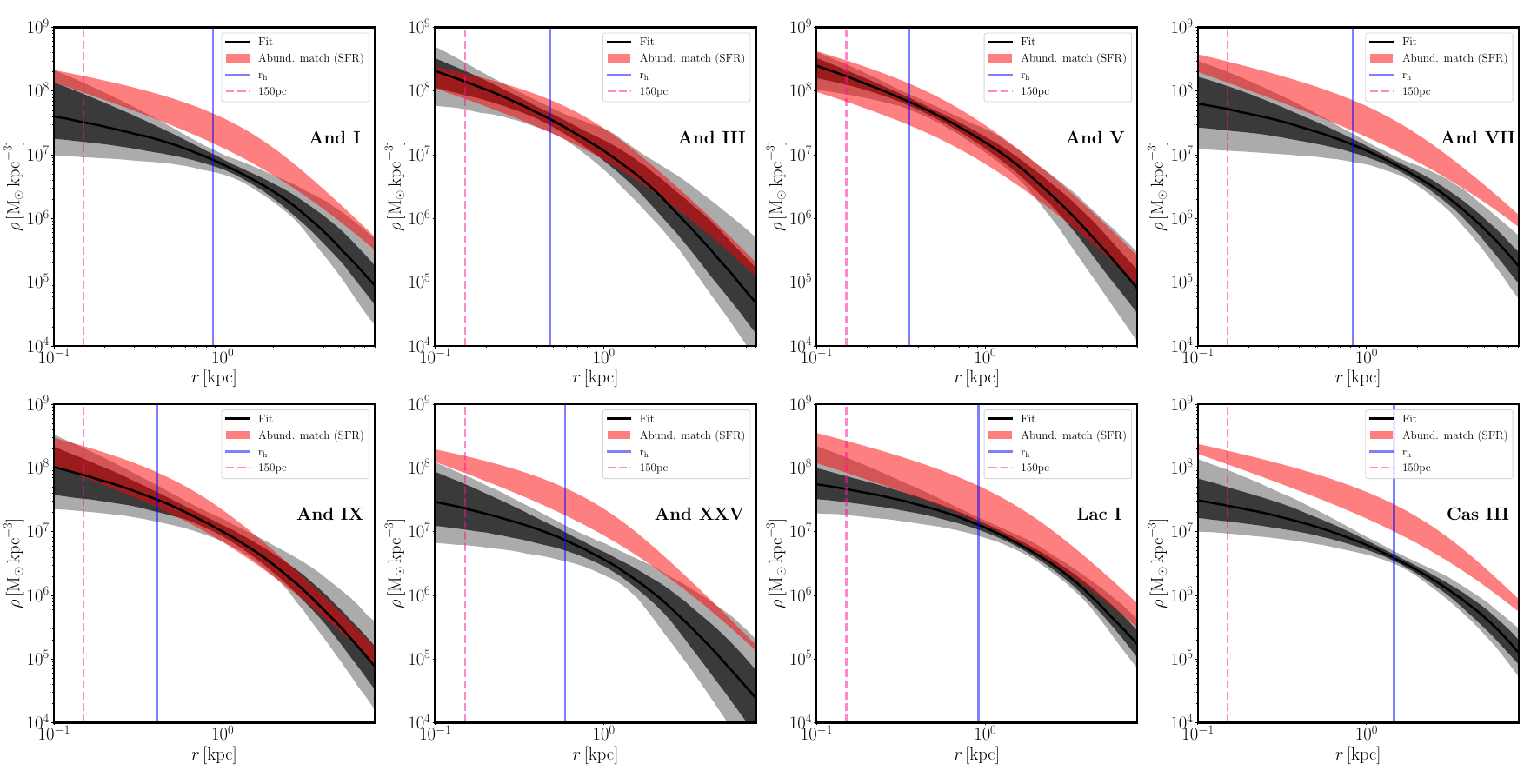}
    \caption{Andromeda dwarf spheroidal dark matter density, $\rho(r)$, profile estimated by \gravsphere ~(solid black line), with $1\sigma$ and $2\sigma$ uncertainties being denoted by the dark and light grey shaded regions, respectively. The red shaded region is the density profile estimated by $\langle{\rm{SFR}}\rangle$ abundance matching, assuming that each dwarf has a median $M_{\rm{200}}-c_{\rm{200}}$ value \citep{ReadErkal_2019}. The solid vertical line is the measured $r_{\rm{h}}$ of each dwarf The dashed vertical line shows 150 pc, the point at which DM density is measured. And I, And VII, And XXV, Lac I, and Cas III all show lower central DM densities at 150 pc and at their respective half-light radii. This is likely caused by a combination of star formation and tidal interactions with M31, which we discuss further in $\S$\ref{subsec:tides}. We note that this this trend could also be explained by scatter in $c_{\rm{200}}$.}
    \label{fig:And_Dwarf_DM_all}
\end{figure*}

\gravsphere ~estimates the dark matter densities of a given system, which are displayed in Figure \ref{fig:And_Dwarf_DM_all}. Each figure shows a solid black line, representing the best-fit DM density, with dark and light grey shaded error bands, representing 1$\sigma$ and 2$\sigma$, respectively. The half-light radius is shown as a solid vertical line, while a distance of 150 pc is denoted by a vertical dashed line. 150 pc is the point at which dark matter density is measured, consistent with \citet{Read_2018}, \citet{Genina_2020}, C21, C23, and Paper I.

\begin{figure}
    \centering
    \includegraphics[width=\columnwidth]{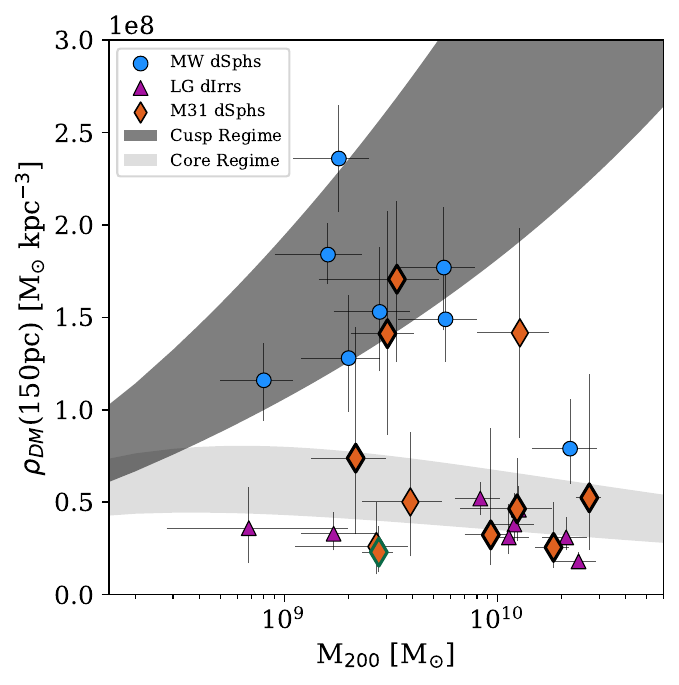}
    \caption{Central dark matter density as a function of pre-infall halo mass, $M_{\rm{200}}$. The dark grey region represents a fully cusped profile, while the light grey region represents a fully cored profile (\texttt{CoreNFW} from \citet{Read_2016}). Both bands of a width corresponding to a 1$\sigma$ scatter in the DM halo concentration \citep{Dutton_2014}. Milky Way dwarf spheroidals are shown as circles, while Local Group dwarf irregulars are shown as triangles. Andromeda dwarf spheroidals are represented by diamonds. Diamonds with a thick border indicate the systems analyzed in this publication. Andromeda XXV is shown with a dark green border, with the only change from \citet{Charles_2022} being an updated $M_{\rm{200}}$ estimate due to new SFH data from \citet{Savino_2023}. Error bars on all points are 1$\sigma$ uncertainties, with the width of And XXV representing its $M_{\rm{200}}$ error. Data for these points were obtained via \citet{Read_2019}. Previous mass-modeled density data for Andromeda dwarfs were taken from C21, C23, and Paper I.}
    \label{fig:DM_summaryplot}
\end{figure}

\begin{figure*}
    \centering
    \includegraphics[width=\linewidth]{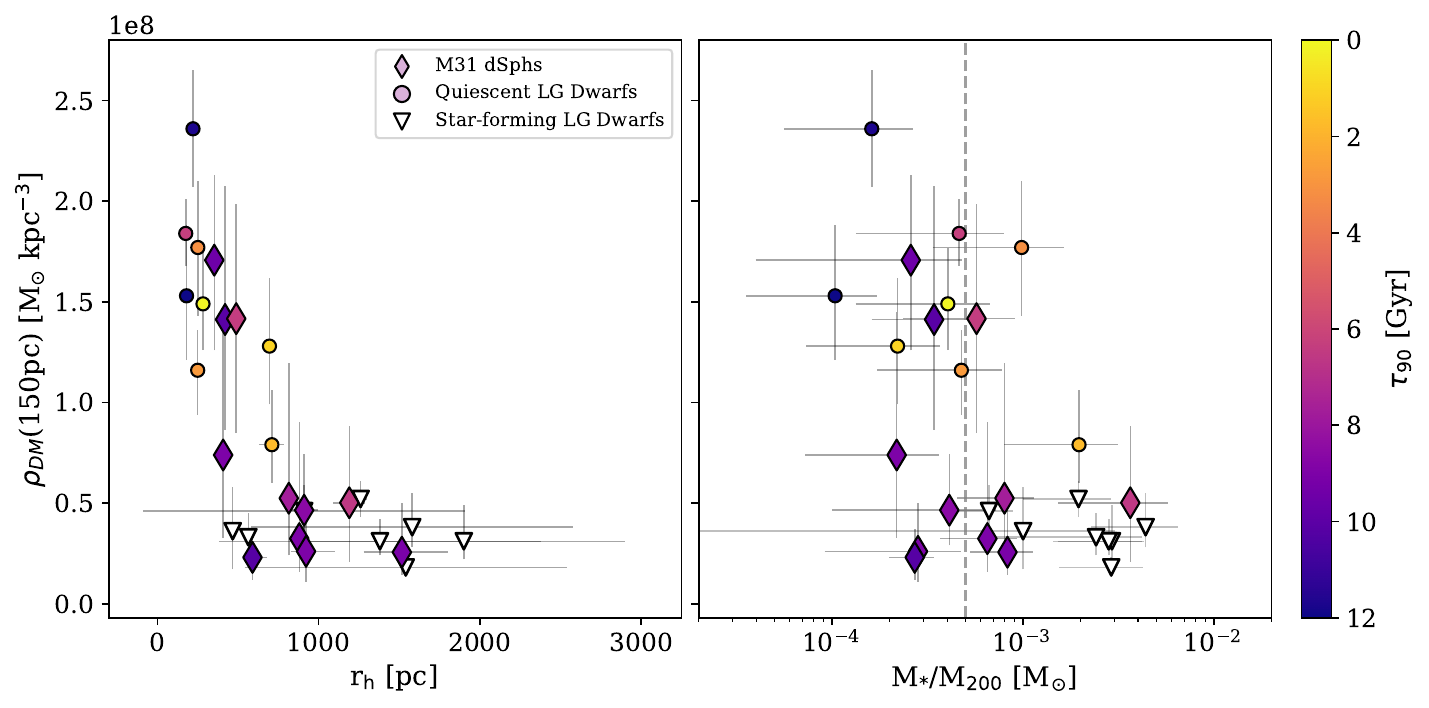}
    \caption{\textbf{Left:} Half-light radius vs. Dark matter density at 150 pc of Local Group dwarfs. At $r_{\rm{h}} \approx 600$ pc, dwarfs begin to appear more cored, largely independent of SFH, with the exception of all star-forming dwarfs falling into the cored regime. \textbf{Right:} DM halo mass vs. Dark matter density at 150 pc of Local Group dwarfs. The dashed vertical line is the threshold below which (leftwards) star formation is too inefficient to lower central DM densities \citep{DiCintio_2014}. Four of the cored M31 dwarfs analyzed in this work lie above this relation, indicating that their SFH could potentially lower their central densities, contrary to Paper I and \citet{Savino_2025}. However, an equal number of cored M31 satellites from this work lie below this threshold, supporting the idea that additional mechanisms may be required to obtain such low central DM densities. This is further supported by the fact that And XXI and And XXV, two  dwarfs predicted to have been significantly affected by tides \citep{Collins_2021, Charles_2022}, also lie below this threshold. \textbf{Both:} Diamonds on both plots represent M31 dwarf satellites, while circles denote quiescent Local Group dwarf galaxies. Upside-down triangles represent star-forming Local Group dwarf systems. M31 and quiescent LG dwarfs are shaded by the point at which they formed 90\% of their stellar mass, $\tau_{\rm{90}}$, as indicated by the color bar on the right ($\tau_{\rm{90}} = 0$ marks present day). All error bars are 1$\sigma$ uncertainty.}
    \label{fig:HM_SF_Comps}
\end{figure*}

The estimated central DM density of each dwarf is shown in Table \ref{tab:dwarf_params}. Our results remain consistent with the idea that the M31 dwarf satellite system contrasts with that of the Milky Way. This trend can be seen clearly in Figure \ref{fig:DM_summaryplot}. Of the seven dwarfs newly analyzed in this paper, two can be considered to have density profiles consistent with an LCDM cusp (hereafter a `cuspy' profile): And III and And V. The remaining five dwarfs analyzed are estimated to have profiles consistent with an LCDM core (hereafter a `cored' profile). Of these cored dwarfs, And I, And VII, Lac I, and Cas III appear to be the most peculiar. Their estimated DM densities from \gravsphere ~fall well below their expected densities from abundance matching (see $\S$\ref{subsubsec:AbundanceMatching}) in \LCDM ~in the absence of any dark matter heating or tides. These abundance matching estimates also result under the assumption that all dwarfs lie on the median $M_{\rm{200}}-c_{\rm{200}}$ relation. These dwarfs appear to mirror star-forming Local Group systems in both central DM density and halo mass, though they have all ceased star formation at least 2 Gyr ago. Furthermore, unlike And XXV, their $M_{\rm{*}}$-$M_{\rm{200}}$ ratio appears to fall above the threshold estimated by \citet{DiCintio_2014}. And XXV falls to the left of this `sweet spot' (shown as the dashed vertical line in the right-hand plot of Figure \ref{fig:HM_SF_Comps}), indicating that its dark matter distribution should not easily be affected by star formation alone. However, the system appears heavily cored. The cored dwarfs from our analysis (And I, And VII, And IX, Lac I, Cas III) lie to the right of this threshold, suggesting that their DM content can become cored from star formation alone. Yet, systems like And V and And III appear rather unaffected by their star formation despite lying to the right of the threshold and despite having similar SFHs as the cored dwarfs. None of these systems show the strong, bursty star formation needed to lower their DM densities, as described by \citet{Pontzen_2012, Pontzen_2014}. These differences and potential causes are further explored in $\S$ \ref{sec:discussion} of this paper.

Despite a multitude of cored systems, we are now able to add two M31 dSphs to the `cuspy' regime of Figure \ref{fig:DM_summaryplot}, with And V notably being the first mass-modeled M31 dwarf to fall most comfortably into this regime. Both And III and And V are now the closest analogues to MW dwarfs, residing well within the neighborhood of quenched MW satellites. Even within 1$\sigma$ uncertainty, And V remains close to its cusped MW counterparts of similar $M_{\rm{200}}$. This not only confirms that \gravsphere ~is able to successfully recover cusped dwarfs at a distance of M31 but also that the Andromeda satellite system is indeed home to higher-central density dwarf galaxies. Furthermore, And III and And V appear to have central densities that remain unaffected by their respective SFHs. Their upper left position in the right-hand plot of Figure \ref{fig:HM_SF_Comps} indicate this, as they are the fall below the quoted efficiency limit (vertical dashed line) of \citet{DiCintio_2014} and retain their DM cusp. 

\subsubsection{Abundance Matching} \label{subsubsec:AbundanceMatching}

Abundance matching is a method in which visible galaxies are mapped to dark matter halos. More specifically, galaxies of a given observed number density can be statistically mapped to simulated DM halos of the same number density (e.g. \citealt{Peacock_2000, Kravtsov_2004, Vale_2004, Vale_2007}). In doing so, a halo mass, $M_{\rm{200}}$, can be derived. In fact, so long as it rises monotonically with halo mass and has negligible scatter, any galactic property can be determined from abundance matching \citep{ReadErkal_2019}. However, using stellar mass, $M_{\rm{*}}$, is limiting for satellite galaxies as they may have had been quenched upon infall into their host or group. This can result in their stellar mass being `frozen in,' and causing scatter below more isolated systems in the in the $M_{\rm{*}}-M_{\rm{200}}$ relation \citep{Geha_2012, Peng_2012, Gatto_2013}. Satellite systems can also appear above the relation if their star formation was shut down by tidal stripping and shocking \citep{Read_2006, Tomozeiu_2016, ReadErkal_2019}. Hence, this relation bias towards isolated systems making $M_{\rm{*}}$ abundance matching alone unreliable for systems such as the M31 dwarfs.

A less biased abundance matching technique uses the mean star formation rate (SFR) of a dwarf, $\langle$SFR$\rangle$, instead of its stellar mass $M_{\rm{*}}$. This value is defined as the SFR averaged over all active star formation times in a galaxy. A full review of this value is shown in $\S$ 4 of \citet{ReadErkal_2019}. Pre-infall halo mass estimates using $\langle$SFR$\rangle$ are shown to be in excellent agreement with pre-infall dynamical $M_{\rm{200}}$ estimates for MW dwarf galaxies. An $\langle \rm{SFR} \rangle-M_{\rm{200}}$ relation also produces less scatter than what is seen in the $M_{\rm{*}}-M_{\rm{200}}$ relation, making it ideal for the analysis of satellite galaxies around systems like the Milky Way and M31. Hence, we use the $\langle$SFR$\rangle$ to determine the halo masses of our M31 dwarfs presented in this work.

As in Paper I, we employ a $\langle$SFR$\rangle$ abundance matching technique to estimate the central DM densities of our dwarfs. We do not apply $M_{\rm{*}}$ abundance matching to our dwarfs in this work, but we direct the interested reader to \citet{ReadErkal_2019} and \citet{Pickett_2025} for such estimations. Our $\langle$SFR$\rangle$ abundance matching process utilizes HST data, presented in \citet{Savino_2025}, for each mass modeled dwarf. These data provide a more complete story of a dwarf's baryonic mass and whether this mass can accurately explain its kinematics today. To compare to the present-day DM densities, we overplot the abundance-matched curves onto the \gravsphere ~results and compare the density at 150 pc. If the abundance-matched densities do not agree with the mass modeled result, we consider this an indication that star formation cannot properly account the contained DM at the central regions. We further take a mismatch of profiles at $r_{\rm{h}}$ to be an indication of tidal effects, which is further explored in $\S$\ref{subsec:tides}. These profile comparisons can be seen in Figure \ref{fig:And_Dwarf_DM_all}, with all of our low-density systems (And I, And VII, Lac I, Cas III) appearing to strongly disagree with their abundance matched profiles beyond 2$\sigma$ at 150 pc and at their respective half-light radii. The potential reasons for this mismatch are further explored in $\S$\ref{sec:discussion}.

\section{Discussion} \label{sec:discussion}

In this section, we discuss key differences in dark matter densities between the Milky Way and M31 populations. We also explore potential causes of these differences, with a focus on Star Formation History, tidal interactions, and different dark matter models. We now have a more complete view of the Andromeda system with which to compare to MW data from \citet{Read_2019}. These comparisons help us to better understand the key factors in the environments of each dSph that may have affected their dark matter content.

It is clear in Figure \ref{fig:DM_summaryplot} that quenched Milky Way dwarfs are more likely to retain high-density dark matter cusps whereas quenched M31 dwarfs are frequently cored. With the exception of three systems, M31 dwarfs fall below their MW counterparts in DM density despite residing in similar-mass halos (based off $M_{\rm{200}}$ estimations). This calls to question how these cores formed.

\subsection{Star formation} \label{subsec:StarFormation}

Early dwarf SFH analyses focused heavily on Milky Way satellites due to their proximity and brightness. As technology progressed, the limited subsample of LG dwarfs (see \citealt{Weisz_2014b}) has since been greatly expanded upon. Through this, it has become clear that MW systems may not be representative of the entire LG. Results from newer studies (i.e. \citealt{Savino_2023, Savino_2025}) have shown that MW satellites typically have less extended SFHs after reionization, forming ${>}50\%$ of their stars $\sim$6-9 Gyr ago. The faintest of these ($M_{\rm{V}} \geq -7$) quenched $\gtrsim$12 Gyr ago, likely due to reionization, with the brightest MW dwarfs all quenching prior to $\sim$1-3 Gyr ago \citep{Weisz_2014b, Weisz_2019}. These trends are not seen in the M31 dwarf population, however, as there appear to be no satellites that quenched 1-3 Gyr ago. While more luminous Andromeda dwarfs rapidly formed $\sim$50\% of their stellar population shortly after reionization, at $\sim$6-9 Gyr ago, many had subsequent delayed quenching times, only forming $\sim$90\% of their stars between 3-6 Gyr ago \citep{Weisz_2019, Savino_2025}. A majority of the M31 dwarf systems, independent of luminosity, also appear to quench almost simultaneously during this time. The dwarfs analyzed in this work are no exception to this trend.

\citet{Savino_2025} discusses that And V appears to have an initial large burst of star formation, with a rapid decline followed by sustained low-level star formation for many Gyrs. This less intense star formation has evidently had little effect on the central DM density of And V, as it is the most cuspy dwarf of our sample so far (see Figure \ref{fig:DM_summaryplot}). Our result for An V therefore aligns with expectations from \citet{Pontzen_2012, Pontzen_2014} in that more sustained, violent star formation is likely required to significantly affect a dwarf's central DM profile. Our other dwarfs also agree with these predictions on the basis of their more extended star formation histories. Works such as \citet{Read_2016} and \citet{Sarrato_2026} show that only a few Gyrs of star formation is sufficient to cause lower-density profiles. The cored dwarfs presented here appear to have more more significant, extended star formation over time compared to And V, which could lead to weaker central cusps. While DM heating via SF can likely explain the inner profiles of the cored dwarfs in this work, four dSphs (And I, And VII, Lac I, Cas III) appear to also have lower densities at their half-light radii. And I lies at a closer deprojected distance to M31 than And V \citep{Savino_2022} and could potentially have been subjected to tidal forces. This is potentially similar to And XXV, which also resides closer to M31 and is predicted to have had significant tidal interactions with its host \citep{Charles_2022}. This is best visualized by the two dwarfs' similarly mismatched DM density profiles to their abundance-matched profiles in Figure \ref{fig:And_Dwarf_DM_all}. Interestingly, And VII, Lac I, and Cas III also present lower densities at both 150 pc and their respective half-light radii, despite being significantly further from their host than And V \citep{Savino_2022}. Conversely, And IX sits at a closer deprojected distance to M31 than And V yet aligns with its expected abundance-matched density at both 150 pc and $r_{\rm{h}}$. The larger uncertainties on the dSph's central density, however, also place it in both the cored and cusped regime more so than its lower-density counterparts. With a half-light radius of $r_{\rm{h}} = 408 \substack{+62 \\ -42}$ pc, it also resides near the transition region of ${\sim}600$ pc for the $\rho_{\rm{DM}}-r_{\rm{h}}$ relation (see Figure \ref{fig:HM_SF_Comps}). Further modeling is required to more definitively understand this `middle-ground' M31 satellite. Regardless, potential external factors are likely needed to explain these density reductions in both regions independent of present-day distance from M31.

As discussed in Paper I, SFHs can reveal significant signatures of changes to a dark matter profile. Extended star formation, or frequent supernova starbursts, can potentially heat up and expel dark matter from a system, leading to the formation of low-density cores \citep{Gnedin_2002, Penarrubia_2012, Pontzen_2012, Pontzen_2014}. In systems where star formation has been shut down, or `quenched,' early in the dwarf's lifetime, and where starbursts are not frequent enough to heat DM, it is expected that density remains unaffected \citep{Moster_2010, Kirby_2011, Brown_2014, Read_2016, Katz_2017, Read_2017, Kravtsov_2018, Read_2018, Charles_2022, Savino_2023, Muni_2025}. Estimating the DM content, in the context of stellar mass, has commonly been performed via the stellar-to-halo mass relation of dwarf galaxies, $M_{\rm{*}}$-$M_{\rm{200}}$ \citep{Penarrubia_2012, DiCintio_2014, deLeo_2024}.  However, there can be significant scatter in this relation, with isolated, quenched systems potentially having their mass `frozen in' (e.g. \citealt{Read_2017, GarrisonKimmel_2017, Nadler_2020, Bouche_2022, Ahvazi_2024}). This has been alleviated by opting to use the mean SFR, $\langle$SFR$\rangle$, as a means of assuming minimal external disruption to an isolated system (see \citealt{ReadErkal_2019}). This method was used to create the density profile comparisons in $\S$\ref{subsubsec:DMdensities} and results in a suitable comparisons for expected vs. estimated DM density at $z\sim0$.

Despite scatter reduction, both \citet{Muni_2025} notes that this relation is not guaranteed to be the sole determinant of a dwarf's DM core size and shape. Through \texttt{EDGE} simulation of faint dwarf galaxies, they introduce a post- and pre-reionization stellar mass ratio, $M_{*, \rm{post}}/M_{*, \rm{pre}}$. This metric accurately correlates with the extent of core formation in DM halos, with a higher value indicating a lower central density and vice-versa. This alternative ratio also correlates with their results showing that the timing of star formation plays a crucial role in DM core formation. Early star formation is shown to relate to higher central densities, while late-time, sustained star formation results in a lower central density. Simulated dwarfs that presented a cuspy profile has formed most of their stars by $z\sim6$, whereas those with cored profiles sustain star formation until $z\sim0$. This links active, post-reionization star formation to lower-density cores. \citet{Sarrato_2026}, using FIRE-2 and Numerical Investigations of Hundred Astrophysical Objects (NIHAO) simulations, explores the proposed relations in \citet{Muni_2025}. They show that burstiness of star formation, as well as its temporal concentration, can drive significant scatter alongside the dwarf's assembly history.

However, this is not the type of star formation seen in our modeled M31 dSphs. Our dwarfs that lie within the cored regime (And I, And VII, Lac I, Cas III) commonly show a disagreement between their expected DM density via $\langle$SFH$\rangle$ and their estimated density from \gravsphere ~(see Figure \ref{fig:And_Dwarf_DM_all}). Meanwhile, the DM densities of the cuspy systems, as well as And IX, are in agreement with their respective $\langle$SFH$\rangle$ estimations at all radii. \citet{Savino_2025} shows the same SFH trend for all dwarfs in our sample, with the majority of star formation occurring before reionization and quenching occurring around the same epoch of ${\sim}5-6$ Gyr ago. While expected densities are being estimated by star formation at 150 pc and the half-light radius for cuspy dwarfs, the question remains: What additional factor is causing the lower DM densities at outer radii of many cored M31 dSphs?

\subsection{Nurture over Nature: the effect of tides} \label{subsec:tides}

An alternative process to dark matter heating that has been commonly explored is tidal interactions. These interactions can be divided into to categories: tidal stripping and tidal shocking. Stripping can decrease a dwarf's dark matter density due to tidal forces affecting a dwarf at all radii, and not just the core. This process can strip away dark matter from the edges of a system, leaving a core of low density DM. It is important to note that, in order for a cusped profile to have its central DM density decrease, $\gtrsim$ 99\% of its mass would need to be stripped away \citep{Penarrubia_2008, Penarrubia_2010, Errani_2018, Errani_2020, Errani_2021, Charles_2022}. However, due to a cored profile being less efficient at retaining its density, less extreme mass loss is needed to have a major effect \citep{Read_2006, Penarrubia_2010, Brooks_2014}. Tidal shocking, on the other hand, is the effect of gravitational fluctuations due to eccentric orbital paths of a dwarf around its host. Dynamical friction can heat up dark matter if the fluctuations occur on a timescale shorter than the dynamical time of the dwarf. The dark matter can then be moved to further orbits, a process that is most effective in an already-cored system. This is described in detail in works such as \citet{Read_2006}, \citet{Errani_2017}, and \citet{vandenBosch_2018}. Tidal shocking is also most effective in dwarfs that reside in a low-concentration halo, as shown by \citet{Amorisco_2019} and \citet{Limberg_2025}.

Tidal effects are likely to cause lower densities seen at the half-light radii in our modeled M31 dwarfs. \citet{Errani_2020}, \citet{Errani_2023}, and \citet{deLeo_2024} find that DM cusps are generally resilient to tides whereas dark matter cores are more susceptible. Tides are unlikely to strongly affect DM within $r_{\rm{h}}$ of dwarf (e.g. \citealt{Hayashi_2003, Read_2006, Kazantzidis_2008, Penarrubia_2008, Kazantzidis_2011, Read_2018}), but if the central region is weakened by SF and the outer regions are tidally stripped, the DM density profile may appear lower than expected at all points \citep{Read_2019}. This seems a likely case for many of our cored M31 dwarfs in this work, as they exhibit low densities at both 150 pc and at their respective half-light radii (see \ref{fig:And_Dwarf_DM_all}). However, it is difficult to draw such a definitive conclusion of tidal stripping without proper motion (PM) measurements. Understanding the orbit of a dwarf satellite around its host can provide insight into its previous interactions, especially one where a dwarf could be tidally shocked at its orbital pericenter \citep{Charles_2022}. However, with a limited number of Andromeda dwarfs having their PMs measured (e.g. \citealt{Casetti_2024, Casetti_2025}), estimating the strength of tidal stripping in our sample is not possible at this time. However, as noted by \cite{Genina_2020}, orbits are not always a strong indicator of host-satellite interaction. Visual indications of such processes, such as tidal tails, are important considerations as well. However, our dwarfs do not appear to have such structures in their photometry. It is possible that our observation depth is too shallow to detect these features and may require follow-up in the future. Kim et al. (in prep.) seeks to apply tidal estimations to Andromeda XXV to estimate the likelihood of tidal interactions and to see if there are any signatures that overlap with other mass modeled M31 dwarfs. Doing so may provide better insight into the history of these systems beyond their SFHs. It will also give clarity on their estimated low DM densities.

\subsection{Alternative Dark Matter Models}

Alternative theories to dark matter, such as Modified Newtonian Gravity (MOND; \citealt{McGaugh_2007, McGaugh_2013}), may be able to explain the low DM densities estimated for M31 dwarfs. Recent investigation into the radial acceleration relation (RAR; \citealt{McGaugh_2016}), which MOND uses to explain the velocity dispersions of dwarf galaxies, has discovered that such a relation may not be universal (see \citealt{Julio_2025} for a review). Scatter below the RAR, due to the `external-field effect,' (EFE; \citealt{McGaugh_2013}) has especially raised concern with MOND. Results of \citet{Julio_2025} show there to be no correlation in the RAR between dwarfs within the EFE and those outside of it. They also note that relativistic modified gravity field theories (i.e. RMOND, \citealt{Skordis_2021}; Aether-Scalar Tensor, \citealt{Durakovic_2024}) better explain the RAR scatter compared to MOND. We showed in Paper I that MOND agrees within 1$\sigma$ with our velocity dispersion estimates for And VI and And XXIII, including EFE estimates. However, this does not hold for other dwarfs in this paper, such as And IX. We measure And IX to have a dispersion of $\sigma_{\rm{IX}} =$ \vdIX ~\kms ~whereas MOND predicts a value of $\sigma_{\rm{EFE, IX}} = 1.5 \substack{+0.6 \\ -0.4}$ \kms \citep{McGaugh_2013}. It is unlikely that MOND can explain the kinematics of dwarfs, especially in the context of the RAR. 

Conversely, alternative dark matter models may better explain the low-density dwarfs around M31. In Paper I, we discussed that processes like gravothermal core-collapse in Self-Interacting Dark Matter \citep{Read_2018, Correa_2021} or weaker cusp formation in Warm Dark Matter \citet{Avila_Reese_2001, Maccio_2012, Shao_2013, Lovell_2014, Schneider_2017} could potentially cause the lower densities we estimate with \gravsphere. With star formation histories failing to completely explain the lower central densities of M31 dwarfs, and with tidal effects being difficult to determine without PM measurements, a cause remains to be determined for our results. Simulated dwarfs around MW and M31 potentials, such as those found in EDGE, can be informative in this regard and may reveal galaxy-host interactions that are not easily directly observed.

\section{Conclusions} \label{sec:conclusions}

In this paper, we expanded upon the analysis and results from \citet{Pickett_2025} by presenting updated kinematics and central DM density estimates for seven more M31 dwarf galaxies. The results were estimated via dynamical mass modeling with \gravsphere ~\textit{v1.5}, consistent with previous analysis. Our results show a continuing trend of lower central densities compared to their MW counterparts, calling to question the potential causes of DM density lowering around M31. Our key findings are below:

\begin{itemize}
    
    \item {We add seven new M31 satellites galaxies to the existing catalog of mass-modeled Local Group dwarfs: Andromeda I, III, V, VII, IX, XXXI, and XXXII. Of these systems, five (I, VII, IX, XXXI, XXXII) reside in the low-density, `cored' regime, with two (III, V) now being added to the higher-density, `cuspy' regime. Overall, this continues the trend from previous works that a majority of M31 dwarf spheroidals have low central dark matter densities. However, this also shows that there are direct MW dSph analogs around M31. The contributing factors to the differences in these populations will be investigated in future works.}
    
    \item{We note that the star formation history of cored M31 dwarfs cannot alone account for their low central densities. While their extended SFHs appear to explain the lower dark matter densities in the central regions of our dwarfs, the lower densities at some systems' respective half-light radii appear to require additional mechanisms. We suspect that tidal interactions have caused dark matter to be removed from the outer radii of the cored M31 dSphs while star formation has lowered the dark matter densities in their central regions. We also suspect that halo concentration ($c_{\rm{200}}$) scatter could present these lower DM densities at the half-light radius. Testing combinations of these effects requires further modeling, ideally in simulations such as EDGE with similar-mass galaxies around M31 and MW gravitational potentials.}

    \item{We propose that the half-light radius of a dwarf galaxy may be indicative of its central dark matter density. Below $\sim600$ pc, dwarfs appear to retain their central dark matter content. Larger systems above this radius begin to present lower central dark matter densities, consistent with dark matter `cores' in the Standard Cosmology. This trend holds for both Milky Way and Andromeda satellites, but a more complete survey is needed to confirm any significant patterns.}

\end{itemize}


\section*{Acknowledgments}

CSP acknowledges support from STFC studentship grant ST/X508810/1. MLMC acknowledges support from STFC grants ST/Y002857/1 and ST/Y002865/1.

CSP acknowledges Dr. Amandine Doliva-Dolinsky, a postdoctoral researcher at the Unviversity of Surrey, for her assistance in retrieving photometric data and for her continued support throughout this project.

\section*{Data Availability}

PAndAS and Subaru Suprime-Cam photometric data are available via the Canadian Astronomy Data Centre (CADC) archive. Pan-STARRS data are available upon reasonable request from Erik Tollerud. Raw spectra obtained with DEIMOS are available via the Keck archive. Fully reduced, one-dimensional spectra and photometry will be made available upon reasonable request to the lead author. Electronic tables with reduced properties (coordinates, magnitudes, velocities, and membership probability) for all stars and identified member stars will be provided on the journal website. Star formation histories were obtained via the MAST HLSP repository, found at \url{https://archive.stsci.edu/hlsp/m31-satellites}. The updated \gravsphere ~\texttt{v1.5} code, with the appropriate \texttt{Binulator} binning method, is available at \url{https://github.com/justinread/gravsphere}. This work has made use of Andrew Pace's Local Volume Database, available at \url{https://github.com/apace7/local_volume_database}
.



\bibliographystyle{mnras}
\bibliography{BibFile}



\appendix 
\section{Supplementary Tables and Diagnostic Plots}
\label{sec:append_plots}

This section contains supplementary tables that contain values used in relevant estimations and plots that further explain relations and fits described throughout this text. Table \ref{tab:prob_priors} shows parameters used to determine the CMD membership probability, as well as SBP constant parameters for each dwarf. These are discussed in Sections \ref{subsec:membership} and \ref{subsec:massprofiles}. Table \ref{tab:vel_priors} shows the MCMC priors used in \texttt{emcee} estimations of systematic velocity and velocity dispersion, as discussed in $\S$\ref{subsubsec:VelCut}. Table \ref{tab:tracer_vals} shows the effective photometric and kinematic tracers used in \texttt{Binulator} analysis, as discussed in $\S$\ref{subsubsec:BinAndGrav}. Table \ref{tab:coreNFWpriors} shows the priors used in the \texttt{CoreNFWTides} model in \gravsphere.

\begin{figure*}
    \includegraphics[width=\textwidth]{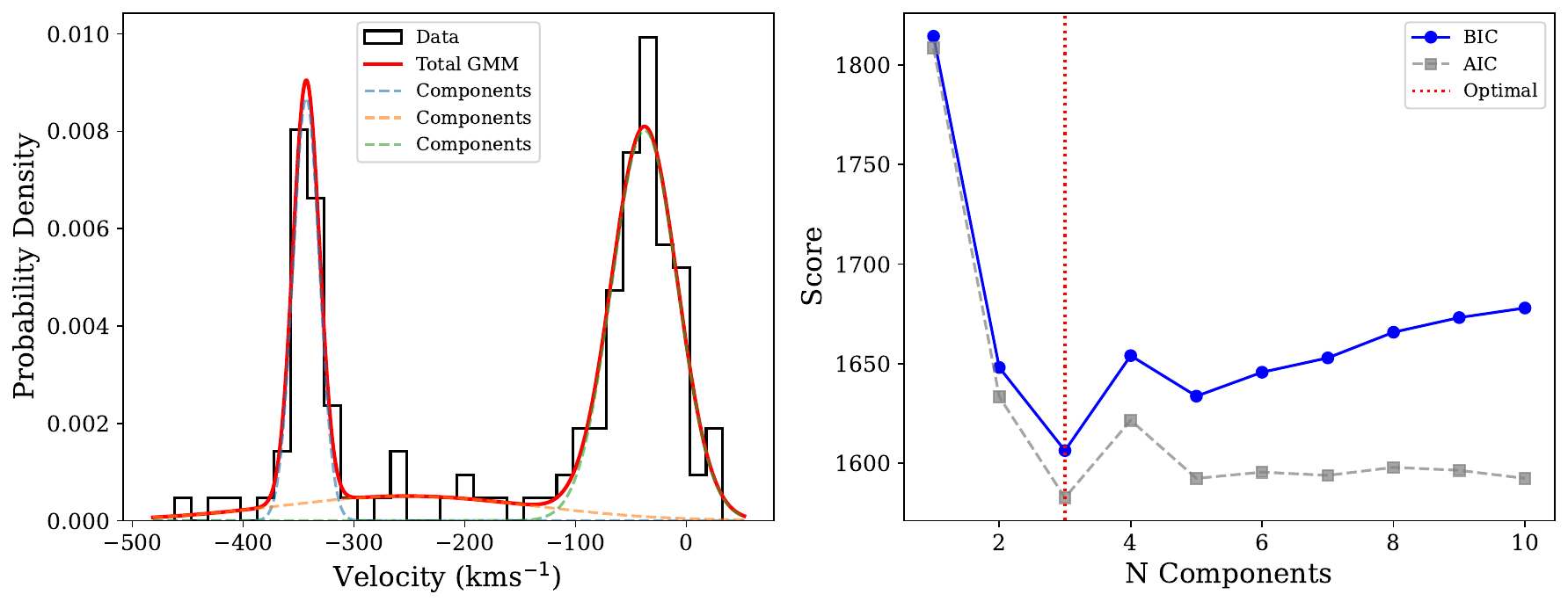}
    \caption{Gaussian Mixture Model test fit of Andromeda III kinematic data. \textbf{Left:} Gaussian fits of the And III velocity distribution, with the dashed lines representing each individual fit applied to the data (represented as a step histogram). The solid line represents the ideal, combined fit determined by our Gaussian Mixture Model. \textbf{Right:} The Bayesian Information Criterion (BIC) results from the Gaussian fits applied in the left-hand plot. The solid blue line represents the BIC, while the long-dashed grey line represents the Akaike Information Criterion (AIC). The short-dashed vertical line shows the point at which the lowest `score' for the BIC and, therefore, the optimal number of fits to our data. For Andromeda III, the ideal number of Gaussian fits is shown to be three.}
    \label{fig:3FitBIC}
\end{figure*}

\begin{figure*}
    \includegraphics[width=\textwidth]{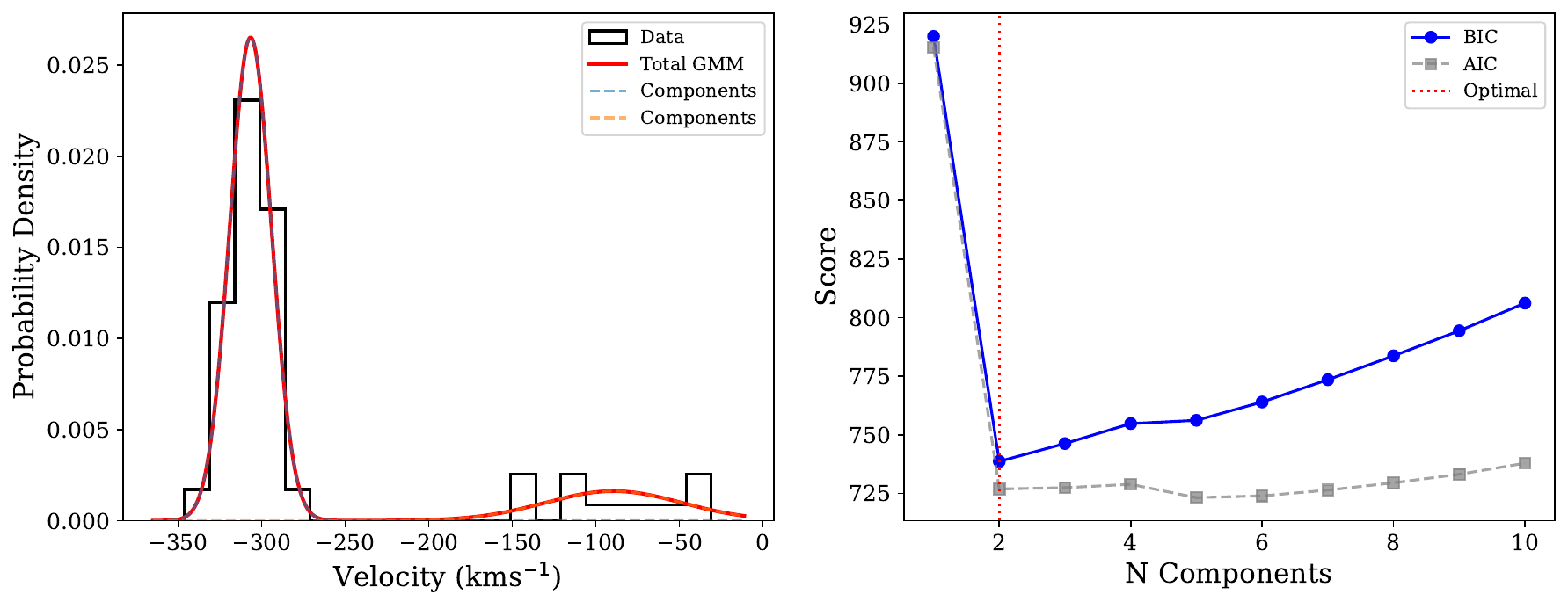}
    \caption{Same as Figure \ref{fig:3FitBIC} but for Andromeda VII. The ideal number of fits for this galaxy is is shown to be two Gaussians, which recovers systemic velocity and velocity dispersion values in agreement with the literature.}
    \label{fig:2FitBIC}
\end{figure*}

Figures \ref{fig:2FitBIC} and \ref{fig:3FitBIC} show Bayesian Information Criterion (BIC) results for And III and And VII, respectively. The purpose of these tests are to avoid over- or under-fitting kinematic data, both of which could result in inaccurate systemic velocity and velocity dispersion values, as noted in $\S$\ref{subsubsec:VelCut}. Figure \ref{fig:2FitBIC} shows a two-Gaussian fit to Andromeda III kinematic data. The left-hand plot shows various Gaussian fits to a velocity histogram, with two distinct peaks (one for the dwarf galaxy and one for foreground Milky Way contamination). The right-hand plot shows the `score' for each fit count, with the lowest BIC score being the optimal number of Gaussians needed to fit the data. For Andromeda III, this value is two. Figure \ref{fig:3FitBIC} is similar but shows a BIC model for And VII, where a single Gaussian is shown to be the optimal fit to the kinematic data.

\begin{figure*}
    \includegraphics[width=\textwidth]{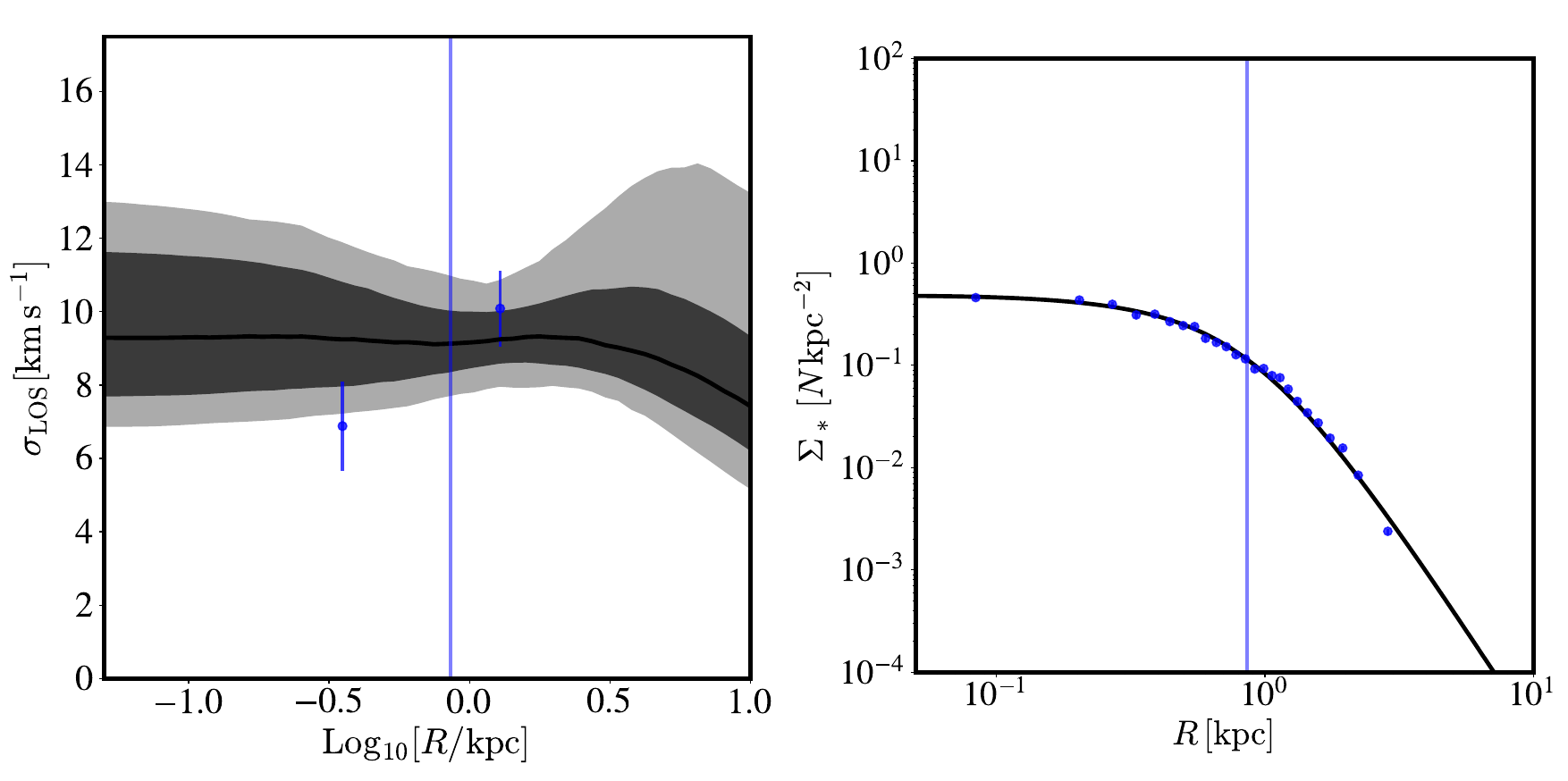}
    \caption{\gravsphere ~diagnostic plots for Andromeda I. \textbf{Left}: Projected line-of-sight velocity dispersion ($\sigma_{\rm{LOS}}$). The dots are binned stellar velocity tracers, as estimated by \texttt{Binulator}, with each bin being shown with an associated error bar. The dark gray contour represents a 68\% confidence interval, while the light gray represents a 95\% confidence interval. \textbf{Right}: The tracer surface density profile ($\Sigma_{*}$). The dots in this plot are the binned stellar photometric tracers, shown with their associated error bars. \textbf{Both}: The solid vertical line represents the half-light radius of the dwarf as calculated by \gravsphere. How these plots are created within \gravsphere ~is explained in \citet{Read_2017}.}
    \label{fig:And1_GSDiag}
\end{figure*}

\begin{figure*}
    \includegraphics[width=\textwidth]{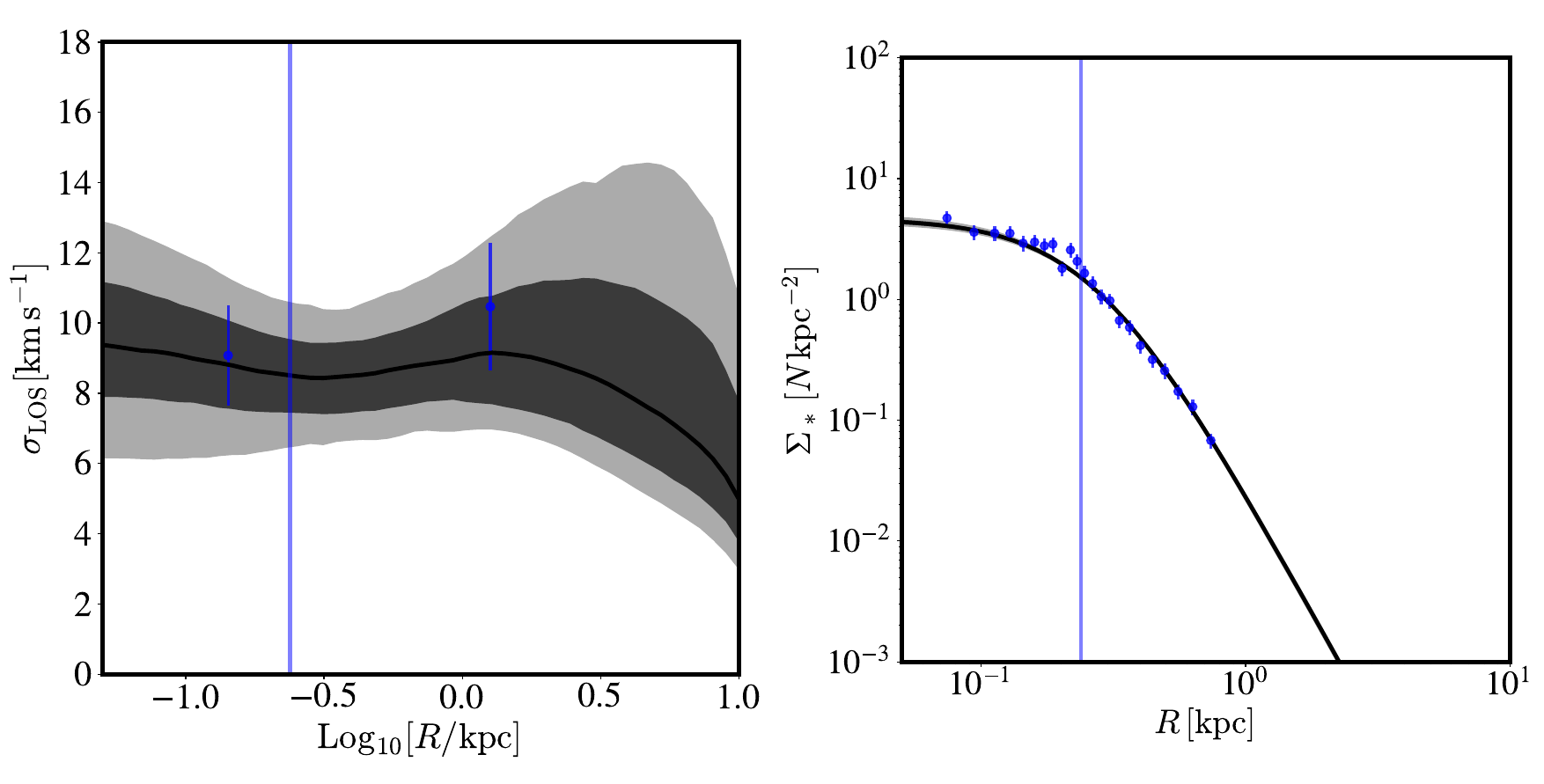}
    \caption{Same as Figure \ref{fig:And1_GSDiag} but for Andromeda III.}
    \label{fig:And3_GSDiag}
\end{figure*}

\begin{figure*}
    \includegraphics[width=\textwidth]{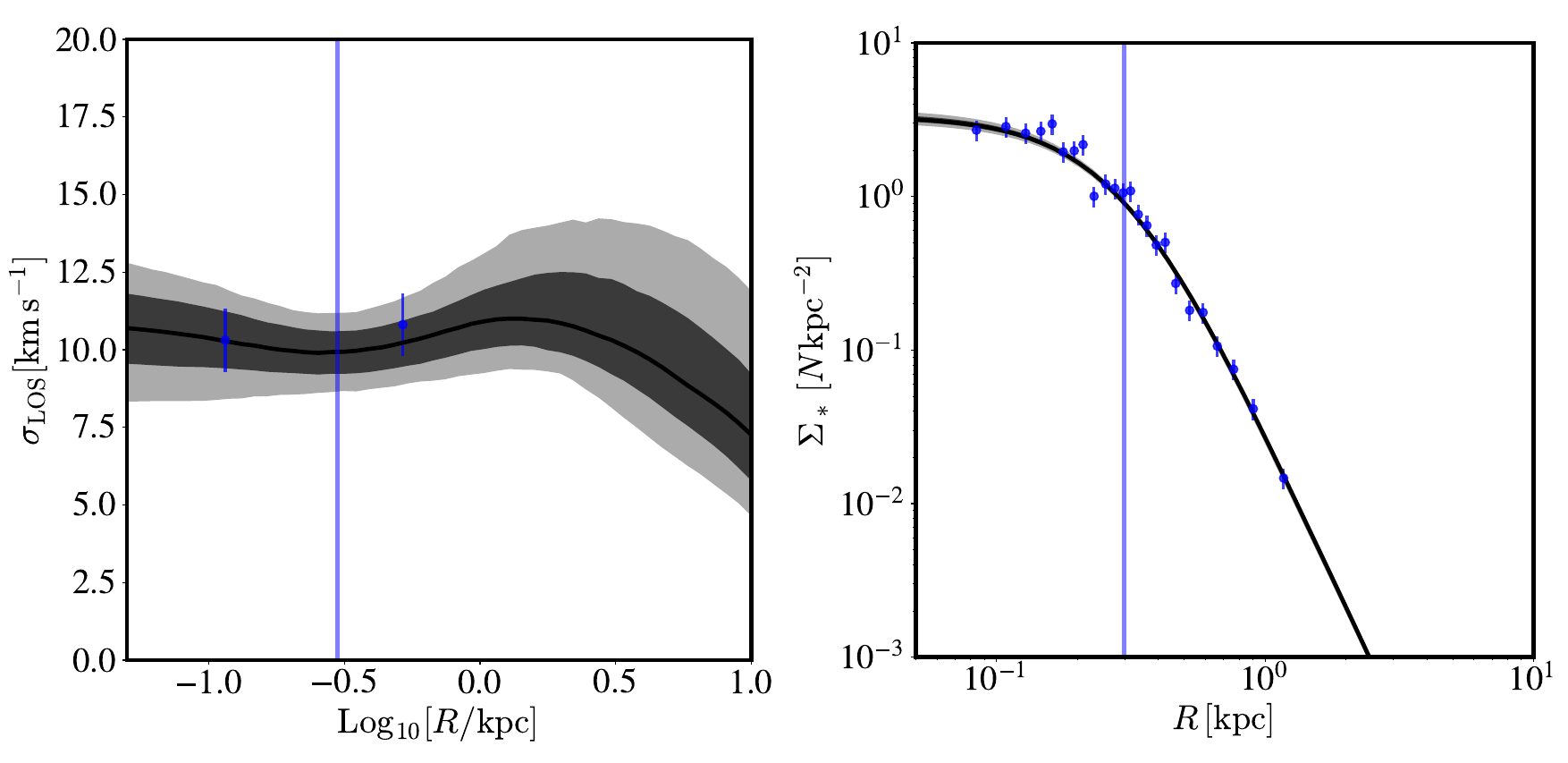}
    \caption{Same as Figure \ref{fig:And1_GSDiag} but for Andromeda V.}
    \label{fig:And5_GSDiag}
\end{figure*}

\begin{figure*}
    \includegraphics[width=\textwidth]{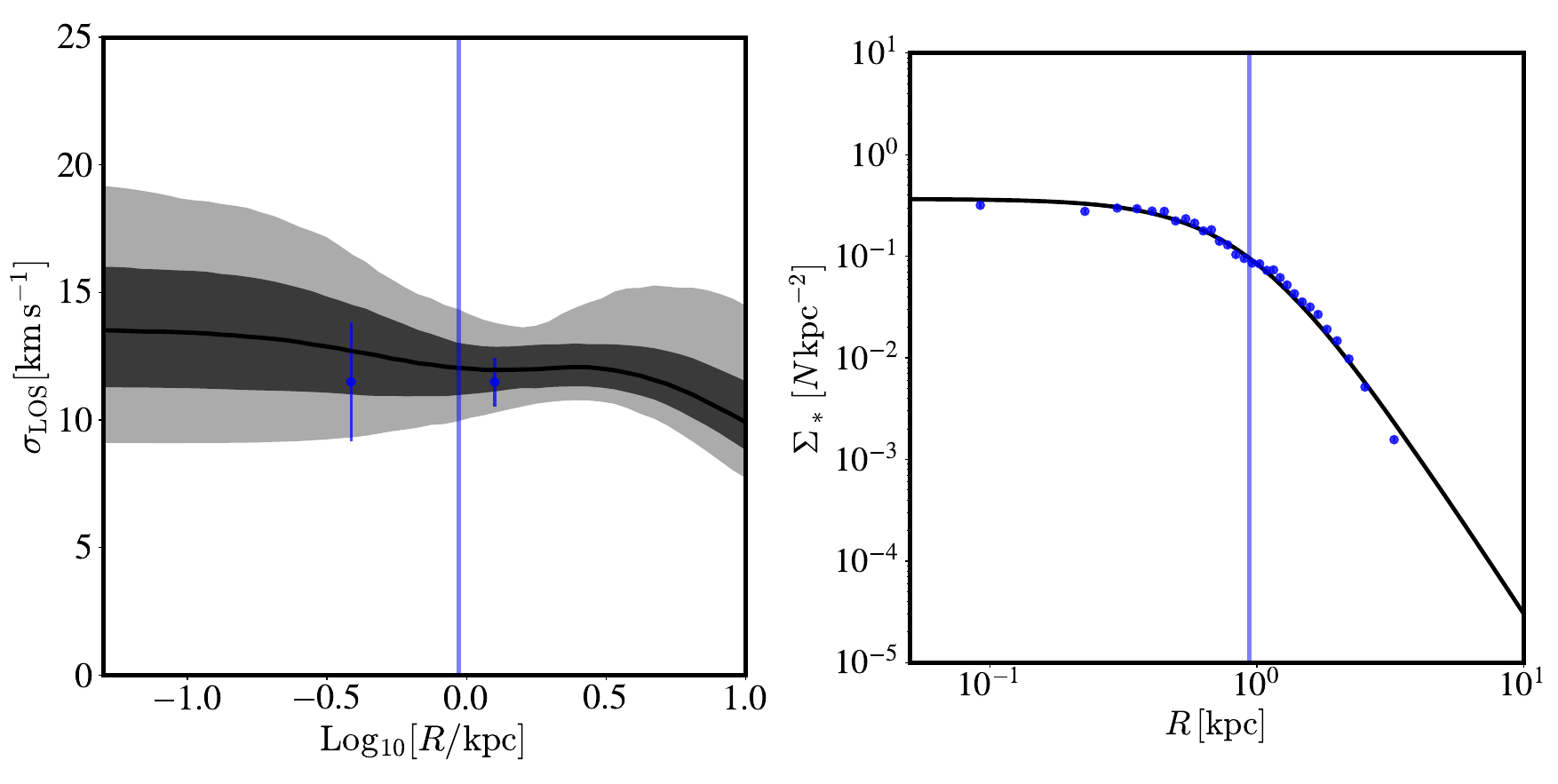}
    \caption{Same as Figure \ref{fig:And1_GSDiag} but for Andromeda VII.}
    \label{fig:And7_GSDiag}
\end{figure*}

\begin{figure*}
    \includegraphics[width=\textwidth]{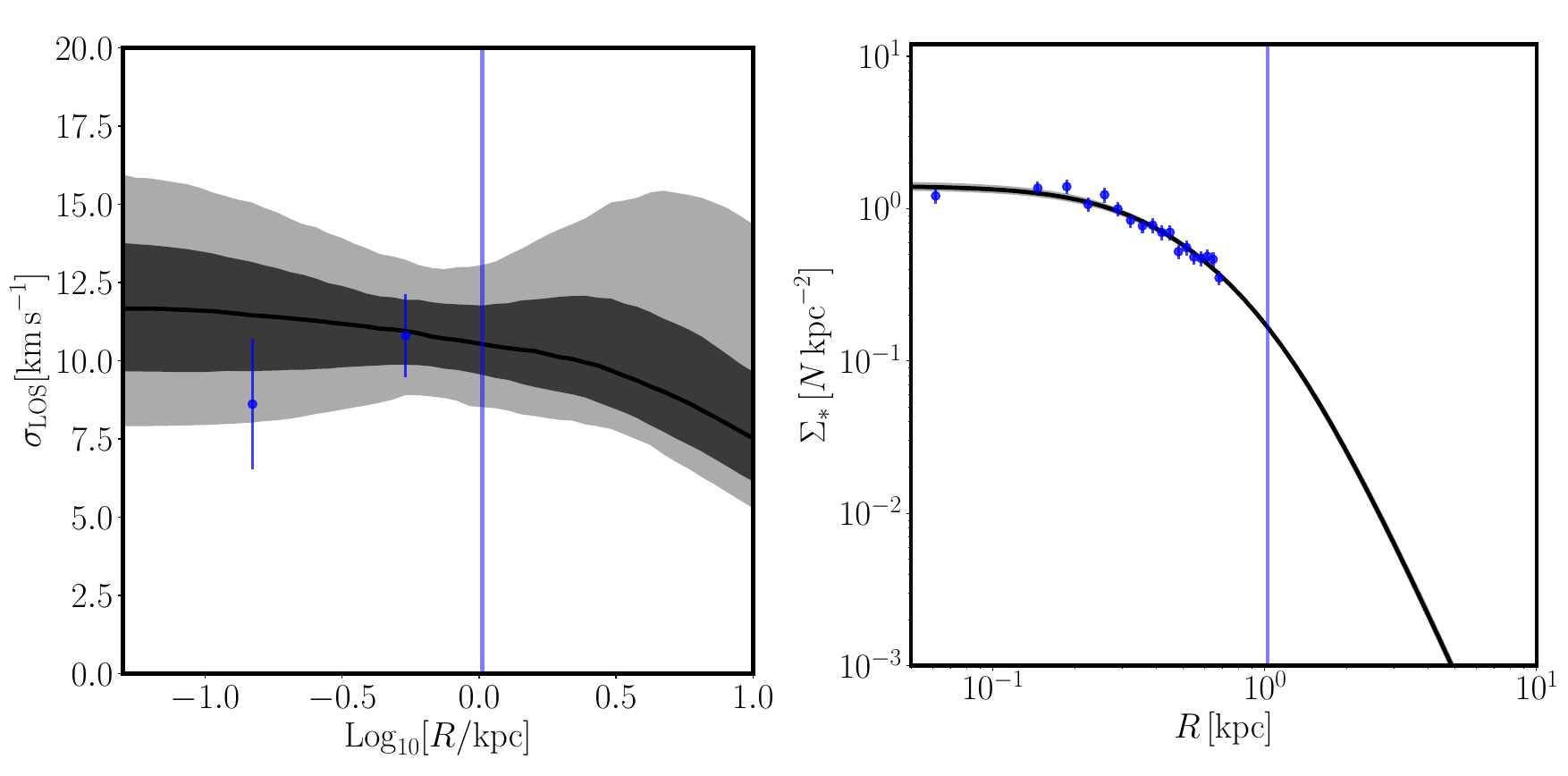}
    \caption{Same as Figure \ref{fig:And1_GSDiag} but for Andromeda IX.}
    \label{fig:And9_GSDiag}
\end{figure*}

\begin{figure*}
    \includegraphics[width=\textwidth]{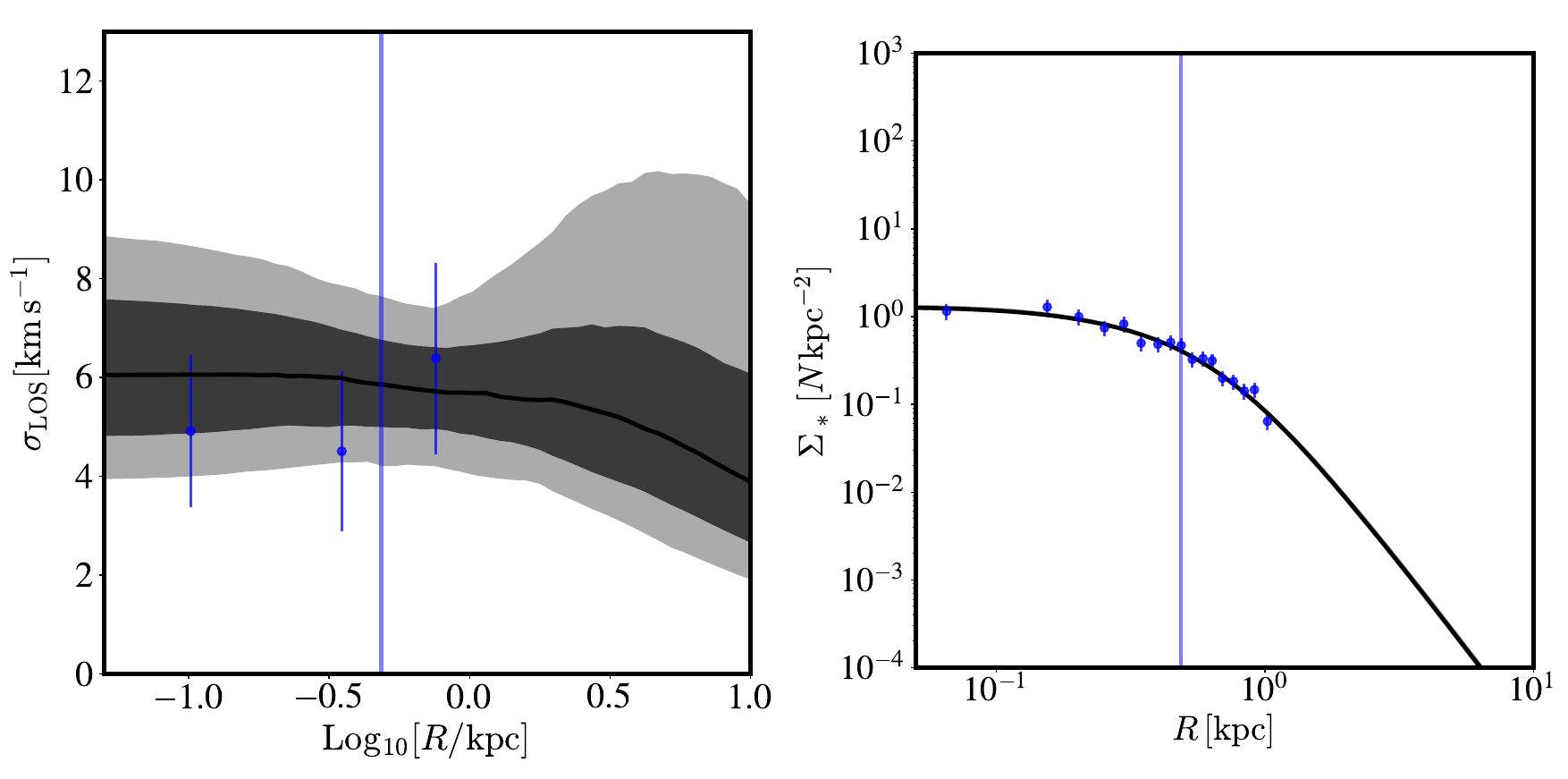}
    \caption{Same as Figure \ref{fig:And1_GSDiag} but for Andromeda XXV.}
    \label{fig:And25_GSDiag}
\end{figure*}

\begin{figure*}
    \includegraphics[width=\textwidth]{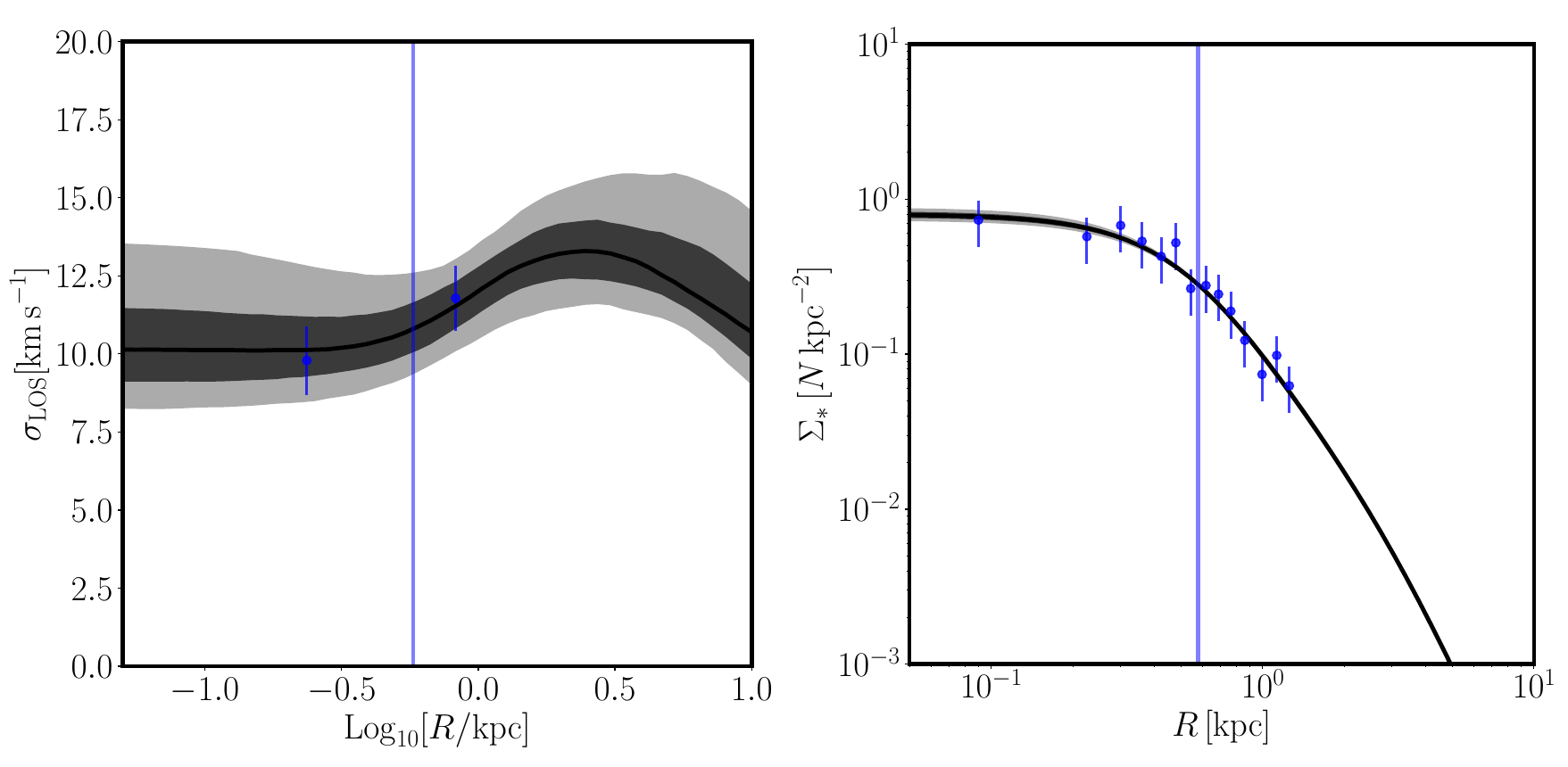}
    \caption{Same as Figure \ref{fig:And1_GSDiag} but for Lacerta I.}
    \label{fig:LacI_GSDiag}
\end{figure*}

\begin{figure*}
    \includegraphics[width=\textwidth]{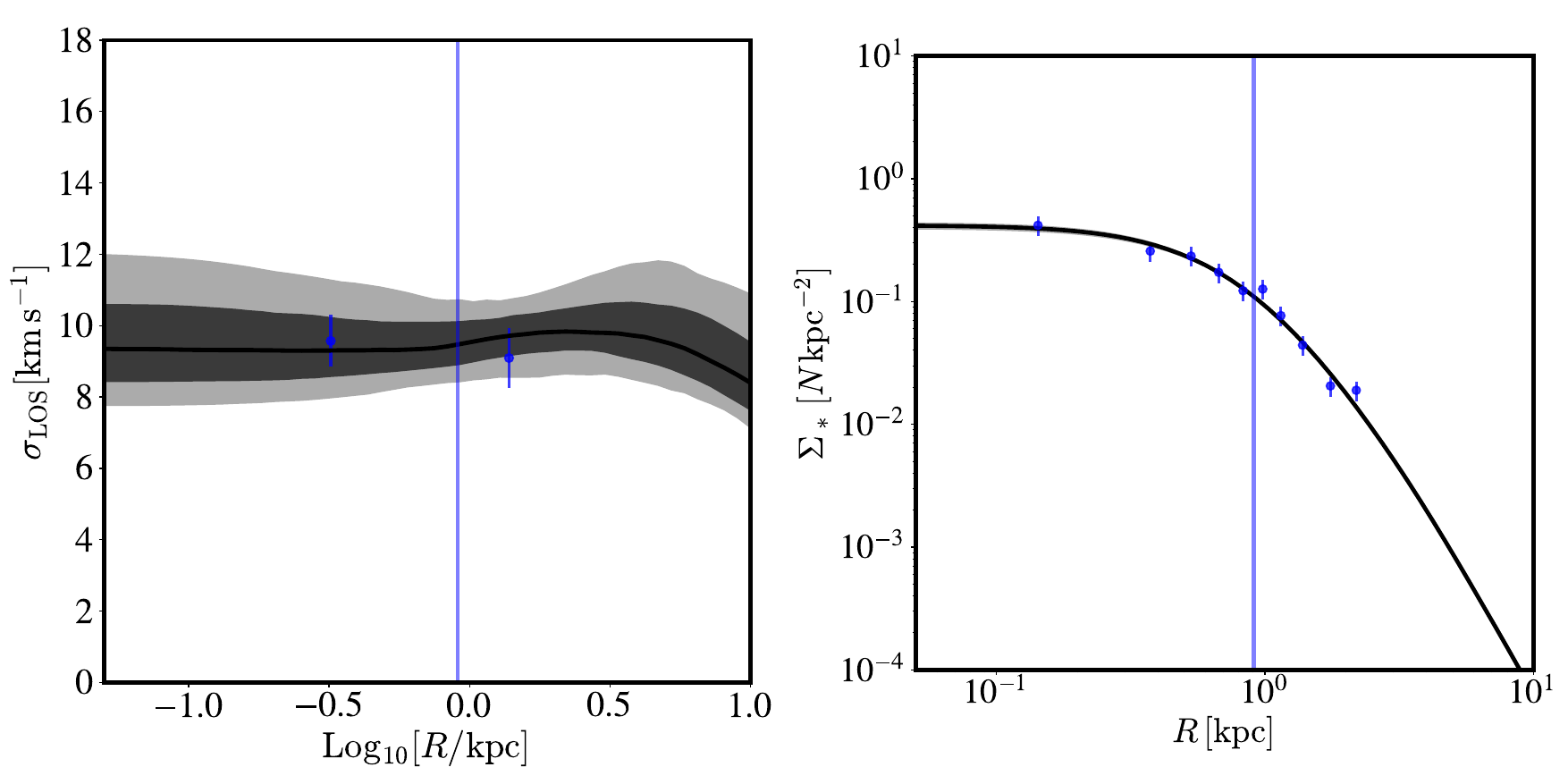}
    \caption{Same as Figure \ref{fig:And1_GSDiag} but for Cassiopeia III.}
    \label{fig:Cas3_GSDiag}
\end{figure*}

Figures \ref{fig:And1_GSDiag} to \ref{fig:Cas3_GSDiag} show \gravsphere ~diagnostic plots for the dwarfs analyzed in this work, as mentioned in $\S$\ref{subsubsec:BinAndGrav}. These figures only show the line-of-sight velocity and tracer surface density profiles, as estimated by \texttt{Binulator} and \gravsphere, in contrast to Paper I. These diagnostics are the most relevant to our dark matter estimations and best depict our constraints on values described throughout this work.


\begin{table}
\renewcommand{\arraystretch}{1.4}
    \caption{CMD membership probability and surface brightness profile constant parameters used for each dwarf. $\eta_{\rm{dist}}$ was rigorously tested through MCMC routines to determine which value best suited our data for a given system. Andromeda XXXI and Andromeda XXXII were not fit with a CMD and therefore so not have CMD probability parameters associated with their analysis. \textit{Note}: The value of $\eta_{\rm{dist}}$ was only used for the creation of a \texttt{Binulator} SBP in this analysis, in contrast with previous analyses where it is also used for a radial distance probability calculation.}
    \label{tab:prob_priors}
    \begin{tabularx}{\columnwidth}{l | Y | Y | Y }
    \hline
        \multicolumn{1}{c}{\textbf{Galaxy}} & \multicolumn{1}{c}{$\eta_{\rm{CMD}}$} & \multicolumn{1}{c}{$\eta_{\rm{CMD, SBP}}$} & \multicolumn{1}{c}{$\eta_{\rm{dist}}$} \\ \hline
        
        $^{a,b}$ And I & 0.2 & 0.4 & 3.5 \\
        
        $^{a,b}$ And III & 0.2 & 0.4 & 4.5 \\   
        
        $^{a,b}$ And V & 0.2 & 0.4 & 3.5 \\
        
        $^{a,b}$ And VII & 0.2 & 0.4 & 4.0 \\
        
        $^{a,b}$ And IX & 0.2 & 0.4 & 4.0 \\
        
        $^{a,b}$ And XXXI & -- & -- & 5.0 \\
        
        $^{a,b}$ And XXXII & -- & -- & 5.0  
\end{tabularx}
\end{table}

\begin{table}
\renewcommand{\arraystretch}{1.4}
    \caption{Priors of the \texttt{emcee} velocity estimations for the dwarf galaxies analyzed in this work. $v_{\rm{r}}$ is the peak velocity and $\sigma_{\rm{r}}$ is the velocity dispersion of a given object. These ranges were informed by quoted literature velocities in: $^{a}$\citet{Tollerud_2012}, $^{b}$\citet{Collins_2013}, and $^{c}$\citet{Martin_2014}.}
    \label{tab:vel_priors}
    \begin{tabularx}{\columnwidth}{l | Y | Y }
    \hline
        \multicolumn{1}{c}{\textbf{Peak}} & \multicolumn{1}{c}{$v_{\rm{r}}$ (kms$^{-1}$)} & \multicolumn{1}{c}{$\sigma_{\rm{r}}$ (kms$^{-1}$)} \\ \hline
        
        $^{a,b}$ P$_{\rm{I}}$   & -400 $ < v_{\rm{r}} < $ -350     & 0 $ < \sigma_{\rm{r}} < $ 20   \\
        
        $^{a,b}$ P$_{\rm{III}}$   & -380 $ < v_{\rm{r}} < $ -310     & "  
                           \\   
        $^{a,b}$ P$_{\rm{V}}$     & -430 $ < v_{\rm{r}} < $ -360     & "                              \\
        $^{a,b}$ P$_{\rm{VII}}$   & -330 $ < v_{\rm{r}} < $ -275     & "                              \\
        $^{a,b}$ P$_{\rm{IX}}$    & -250 $ < v_{\rm{r}} < $ -175     & "                              \\
        $^{c}$ P$_{\rm{XXXI}}$  & -250 $ < v_{\rm{r}} < $ -150     & "                              \\
        $^{c}$ P$_{\rm{XXXII}}$ & -410 $ < v_{\rm{r}} < $ -330     & " 
\end{tabularx}
\end{table}

\begin{table}
\renewcommand{\arraystretch}{1.4}
    \caption{Effective photometric and kinematic tracers, as well as their associated bins, used in \texttt{Binulator}. The low number of kinematic tracers for And XXXI and And XXXII can be attributed to a cut performed on the surface brightness profile to minimized fitting of background data. \textit{Note:} N$_{\rm{bin}}$ represents the number of \textit{stars} in a given bin, not the total number of bins themselves.}
    \label{tab:tracer_vals}
    \begin{tabularx}{\columnwidth}{l | Y | Y | Y | Y }
    \hline
        \multicolumn{1}{c}{\textbf{Galaxy}} & \multicolumn{1}{c}{N$_{\rm{eff,kin}}$} & \multicolumn{1}{c}{N$_{\rm{bin,kin}}$} & \multicolumn{1}{c}{N$_{\rm{eff,phot}}$} & \multicolumn{1}{c}{N$_{\rm{bin,phot}}$} \\ \hline
        
        $^{a,b}$ And I & 27 & 13 & 7193 & 287 \\
        
        $^{a,b}$ And III & 29 & 14 & 1230 & 49 \\  
        
        $^{a,b}$ And V & 53 & 26 & 1063 & 42 \\
        
        $^{a,b}$ And VII & 54 & 26 & 8208 & 273 \\
        
        $^{a,b}$ And IX & 17 & 8 & 3877 & 193 \\
        
        $^{a,b}$ And XXXI & 79 & 39 & 148 & 9 \\
        
        $^{a,b}$ And XXXII & 135 & 78 & 315 & 32
\end{tabularx}
\end{table}

\newpage
\begin{sidewaystable*}
    \vspace*{-18cm}
    \setlength{\tabcolsep}{14pt}
    \renewcommand{\arraystretch}{1.7}
    \centering

    \caption{\textbf{Top:} Priors used for the \texttt{CoreNFWTides} models in \gravsphere ~for each dwarf. All parameters represent the same values as in Paper I: $M_{\rm{200}}$ is the enclosed mass at the virial radius. $c_{\rm{200}}$ is the dimensionless concentration parameter. $r_{\rm{c}}$ represents the size of a constant density core, and $r_{\rm{t}}$ represents the outer `tidal radius' at which density fall-off steepens at a rate of $\rho \propto r^{-\delta}$. $n$ sets the shape of the dark matter density profile, with $n = 0$ corresponding to a central constant density core and $n = 1$ representing a cusp of $\rho \propto r^{-1}$. \textbf{Bottom:} Velocity anisotropy priors used for each dwarf in \gravsphere. $\tilde{\beta}_{\rm{r_{0}}}$ is a transition radius, with $\tilde{\beta}_{\rm{n}}$ controlling the transition sharpness. $\tilde\beta_{\rm{0}}$ is the central symmetrized velocity anisotropy profile. $\tilde\beta_{\rm{\infty}}$ is the outer symmetrized velocity anisotropy profile ($\tilde\beta = -1$ represents a fully tangential distribution, $\tilde\beta = 1$ represents a fully radial distribution. See \citet{Read_2018}, \citet{Collins_2021}, and Paper I for further details of \texttt{CoreNFWTides} and velocity anisotropy parameters.}
    \label{tab:coreNFWpriors}
    \begin{tabular}{l c c c c c c}
        \toprule
        \textbf{Galaxy} &  $\log_{10} (\frac{M_{\rm{200}}}{M_{\odot}})$ & $c_{200}$ & $\log_{10}(\frac{r_{\rm{c}}}{\text{kpc}})$ & $\log_{10}(\frac{r_{\rm{t}}}{\text{kpc}})$ & $\delta$ & $n$ \\ \hline
        
        And I & $9.0 < \log_{10} (\frac{M_{\rm{200}}}{M_{\odot}}) < 11.0$ & $8.4 < c_{200} < 37.1 $ & $0.01 < \log_{10}(\frac{r_{\rm{c}}}{\text{kpc}}) < 10.0$ & $1.0 < \log_{10}(\frac{r_{\rm{t}}}{\text{kpc}}) < 20.0$ & $3.01 < \delta < 4.0$ & $0.0 < n < 1.0$ \\
        
        And III & $8.0 < \log_{10} (\frac{M_{\rm{200}}}{M_{\odot}}) < 11.0$ & $8.4 < c_{200} < 46.9$ & $0.01 < \log_{10}(\frac{r_{\rm{c}}}{\text{kpc}}) < 10.0$ & $1.0 < \log_{10}(\frac{r_{\rm{t}}}{\text{kpc}}) < 20.0$ & $3.01 < \delta < 4.0$ & $0.0 < n < 1.0$ \\
        
        And V & $9.0 < \log_{10} (\frac{M_{\rm{200}}}{M_{\odot}}) < 10.0$ & $10.5 < c_{200} < 37.1$ & $0.01 < \log_{10}(\frac{r_{\rm{c}}}{\text{kpc}}) < 10.0$ & $1.0 < \log_{10}(\frac{r_{\rm{t}}}{\text{kpc}}) < 20.0$ & $3.01 < \delta < 4.0$ & $0.0 < n < 1.0$ \\
        
        And VII & $9.0 < \log_{10} (\frac{M_{\rm{200}}}{M_{\odot}}) < 11.0$ & $8.4 < c_{200} < 37.1$ & $0.01 < \log_{10}(\frac{r_{\rm{c}}}{\text{kpc}}) < 10.0$ & $1.0 < \log_{10}(\frac{r_{\rm{t}}}{\text{kpc}}) < 20.0$ & $3.01 < \delta < 4.0$ & $0.0 < n < 1.0$\\
    
        And IX & $8.0 < \log_{10} (\frac{M_{\rm{200}}}{M_{\odot}}) < 10.5$ & $9.4 < c_{200} < 46.9$ & $0.01 < \log_{10}(\frac{r_{\rm{c}}}{\text{kpc}}) < 10.0$ & $1.0 < \log_{10}(\frac{r_{\rm{t}}}{\text{kpc}}) < 20.0$ & $3.01 < \delta < 4.0$ & $0.0 < n < 1.0$ \\

        And XXV & $8.5 < \log_{10} (\frac{M_{\rm{200}}}{M_{\odot}}) < 10.5$ & $9.4 < c_{200} < 41.7$ & $0.01 < \log_{10}(\frac{r_{\rm{c}}}{\text{kpc}}) < 10.0$ & $1.0 < \log_{10}(\frac{r_{\rm{t}}}{\text{kpc}}) < 20.0$ & $3.01 < \delta < 4.0$ & $0.0 < n < 1.0$ \\
        
        And XXXI & $9.0 < \log_{10} (\frac{M_{\rm{200}}}{M_{\odot}}) < 10.0$ & $10.5 < c_{200} < 37.1$ & $0.01 < \log_{10}(\frac{r_{\rm{c}}}{\text{kpc}}) < 8.0$ & $1.0 < \log_{10}(\frac{r_{\rm{t}}}{\text{kpc}}) < 20.0$ & $3.01 < \delta < 4.0$ & $0.0 < n < 1.0$ \\
        
        And XXXII & $9.0 < \log_{10} (\frac{M_{\rm{200}}}{M_{\odot}}) < 10.0$ & $10.5 < c_{200} < 37.1$ & $0.01 < \log_{10}(\frac{r_{\rm{c}}}{\text{kpc}}) < 10.0$ & $1.0 < \log_{10}(\frac{r_{\rm{t}}}{\text{kpc}}) < 20.0$ & $3.01 < \delta < 4.0$ & $0.0 < n < 1.0$ \\
        \bottomrule
    \end{tabular}

    \vspace*{1.0cm}
        \renewcommand{\arraystretch}{1.75}
        \setlength{\tabcolsep}{40pt}
        \begin{tabular}{l c c c c c} 
        \toprule
        \textbf{Galaxy} & $\tilde{\beta}_{\rm{r_{0}}}$ & $\tilde{\beta}_{\rm{n}}$ &  $\tilde{\beta}_{\rm{0}}$ & $\tilde{\beta}_{\infty}$  \\ \hline
        
        And I & $-2.0 < \tilde{\beta}_{\rm{r_{0}}} < 0.0$ & $1.0 < \tilde{\beta}_{\rm{n}} < 3.0$ & $-0.01 < \tilde{\beta}_{\rm{0}} < 0.01$ & $-0.1 < \tilde{\beta}_{\rm{\infty}} < 1.0$ \\
        
        And III & $-2.0 < \tilde{\beta}_{\rm{r_{0}}} < 0.0$ & $1.0 < \tilde{\beta}_{\rm{n}} < 3.0$ & $-0.01 < \tilde{\beta}_{\rm{0}} < 0.01$ & $-0.1 < \tilde{\beta}_{\rm{\infty}} < 1.0$ \\
        
        And V & $-2.0 < \tilde{\beta}_{\rm{r_{0}}} < 0.0$ & $1.0 < \tilde{\beta}_{\rm{n}} < 3.0$ & $-0.01 < \tilde{\beta}_{\rm{0}} < 0.01$ & $-0.1 < \tilde{\beta}_{\rm{\infty}} < 1.0$ \\
        
        And VII & $-2.0 < \tilde{\beta}_{\rm{r_{0}}} < 0.0$ & $1.0 < \tilde{\beta}_{\rm{n}} < 3.0$ & $-0.01 < \tilde{\beta}_{\rm{0}} < 0.01$ & $-0.1 < \tilde{\beta}_{\rm{\infty}} < 1.0$ \\
    
        And IX & $-2.0 < \tilde{\beta}_{\rm{r_{0}}} < 0.0$ & $1.0 < \tilde{\beta}_{\rm{n}} < 3.0$ & $-0.01 < \tilde{\beta}_{\rm{0}} < 0.01$ & $-0.1 < \tilde{\beta}_{\rm{\infty}} < 1.0$ \\

        And XXV & $-2.0 < \tilde{\beta}_{\rm{r_{0}}} < 0.0$ & $1.0 < \tilde{\beta}_{\rm{n}} < 3.0$ & $-0.01 < \tilde{\beta}_{\rm{0}} < 0.01$ & $-0.1 < \tilde{\beta}_{\rm{\infty}} < 1.0$ \\
        
        And XXXI & $-2.0 < \tilde{\beta}_{\rm{r_{0}}} < 0.0$ & $1.0 < \tilde{\beta}_{\rm{n}} < 3.0$ & $-0.01 < \tilde{\beta}_{\rm{0}} < 0.01$ & $-0.1 < \tilde{\beta}_{\rm{\infty}} < 1.0$ \\
        
        And XXXII & $-2.0 < \tilde{\beta}_{\rm{r_{0}}} < 0.0$ & $1.0 < \tilde{\beta}_{\rm{n}} < 3.0$ & $-0.01 < \tilde{\beta}_{\rm{0}} < 0.01$ & $-0.1 < \tilde{\beta}_{\rm{\infty}} < 1.0$ \\
        \bottomrule
\end{tabular}
\vspace*{\fill}
\end{sidewaystable*}

\label{lastpage}
\bsp	
\end{document}